\documentclass[12pt,preprint]{aastex}

\def\apj{{ApJ}}

\received{}
\accepted{}
\shorttitle{QUEST --- X-ray Perspective}
\shortauthors{Teng et al.}
\journalid{}{}
\articleid{}{}

\begin{document}

\title{X-QUEST: A Comprehensive X-ray Study of Local ULIRGs and QSOs}

\author{Stacy H. Teng \altaffilmark{1,2} and Sylvain Veilleux \altaffilmark{2}}

\altaffiltext{1}{Contacting author: stacyt@astro.umd.edu.}
\altaffiltext{2}{Department of Astronomy, University of Maryland,
  College Park, MD 20742, U.S.A.}

\begin{abstract}

  We present results from the X-ray portion of a multi-wavelength
  study of local ULIRGs and QSOs called QUEST (Quasar-ULIRG Evolution
  STudy).  The data consist of new and archival X-ray data on 40
  ULIRGs and 26 PG QSOs taken with {\em Chandra} and {\em
    XMM-Newton}. A combination of traditional and hardness ratio
  spectral fitting methods is used to characterize the X-ray
  properties of these objects.  
  The absorption-corrected 2-10 keV to bolometric luminosity
  ratios of the ULIRGs and PG~QSOs suggest that the likelihood for
  dominant nuclear activity increases along the merger sequence from
  ``cool'' ULIRGs, ``warm'' ULIRGs, infrared-bright QSOs, and
  infrared-faint QSOs. The starburst dominates the total power in
  ULIRGs prior to the merger, and this is followed by rapid black hole
  growth during and after coalescence. These results are in general
  agreement with those obtained in the mid-infrared with {\em Spitzer}
  and recent numerical simulations.

\end{abstract}


\section{Introduction}
\label{sec:intro}

Since their wide-spread detection by the \textit{IRAS} satellite over
20 years ago, the energy source in luminous ($L_{IR} \geq
10^{11}~L_\odot$) and ultraluminous ($L_{IR} \geq 10^{12}~L_\odot$)
infrared galaxies (U/LIRGs) has been under debate.  It is thought that
reprocessed starlight and/or active galactic nuclei (AGNs) are
responsible for the enormous luminosities these objects exhibit in the
infrared, but the exact contribution from each component is hard to
determine.  \citet{sanders88} proposed that U/LIRGs and quasars are linked through galaxy mergers and ULIRGs are simply the dust-enshrouded phase in the merging process.  The authors suggested that the energy source evolves along the merger sequence from starburst-dominated LIRGs to AGN-dominated ULIRGs and ultimately quasars.  The question of the energy source in U/LIRGs has cosmological implications.  At higher redshifts ($z \gtrsim 1$), U/LIRGs are a significant population, contributing in large part to the cosmic star formation \citep[e.g.,][]{sanmir, blain}.  About a third of the high-$z$ ULIRGs are ongoing gas-rich mergers \citep[e.g.,][]{daddi, shapiro, forster}.  The fraction of high-$z$ ULIRGs which are mergers increases with luminosity \citep[e.g.,][]{tacconi}.  Merger simulations by \citet{li07} suggest that vigorous star formation is needed to reproduce the observed dust properties of high redshift quasars.  \citet{li08} were able to reproduce the observed spectral energy distribution of SDSS~J1148+5251, a quasar at $z \sim 6.4$ using these merger simulations.  Their work support the starburst-to-quasar evolutionary scenario for some luminous quasars at high redshifts.  However, at high redshifts, secular processes may play a more dominant role than in the local universe and the merger scenario may not be necessary for the formation of some quasars \citep[e.g.,][]{dekel}.  If the generalized merger scenario holds true, then a large fraction of the high-$z$ ULIRGs may be the progenitors of some of the present-day quasars.  

Higher redshift objects are difficult to study, so as a first step in understanding the role U/LIRGs play in the formation of some quasars and elliptical galaxies, we utilize local U/LIRGs and quasars as laboratories for studying merger physics. 
Past ground-based optical
and near-infrared photometry have shown that essentially all ULIRGs
show signs of interaction, indicating that they are ongoing mergers
\citep[e.g.,][]{kim02, vei02}. Recent simulations support the scenario
that quasars can indeed be formed through gas-rich galaxy mergers
\citep[e.g.,][]{hopkins05, hopkins06, hopkins08}.  Therefore, the two populations may be linked.

To further study this relationship, we are conducting a comprehensive,
multi-wavelength imaging and spectroscopic survey of local ULIRG and
QSO mergers called QUEST --- Quasar/ULIRG Evolution STudy.  The goal of the project is to investigate the evolutionary link between the two groups of luminous objects in the nearby universe.  The sample
consists of $z < 0.3$ 1-Jy ULIRGs and Palomar Green (PG) quasars.  The 1-Jy sample of ULIRGs is a group of very well studied objects in the infrared and optical \citep[e.g.,][]{vei02, kim02}.  We pair the ULIRGs with another group of objects with a wealth of ground- and space-based data --- the PG~QSOs.  At luminosities above $10^{12}$~L$_\odot$, ULIRGs and PG~QSOs are the only two types of extragalactic sources with similar space densities and bolometric luminosities in the local universe \citep{sanmir}.  While in some respect the PG~QSOs are not representative of all quasars as a group, \citet{jester} have shown that for PG~QSOs with redshifts below 0.5, the selection biases are minimal.  The
detailed QUEST sample selection is described in \citet{vei09a} (hereafter V09a) and
references therein. 

To date, optical and infrared portions of QUEST have found that ULIRGs and PG~QSOs have similar galactic structure and stellar host dynamics.  Results from an {\it HST} NICMOS imaging study by \citet{vei06, vei09b} suggest that AGN-like objects, including the QSOs, are generally of early morphological type and have less pronounced merger-induced morphological anomalies than systems with LINER-like or HII region-like spectral types and cooler infrared colors ($f_{25}/f_{60} \le$ 0.2). Infrared-bright QSOs generally have more pronounced merger-induced morphological anomalies than infrared-faint QSOs.  The velocity dispersion distributions in ULIRGs resemble those in intermediate mass ellipicals/lenticulars with moderate rotation and the black hole masses of these ULIRGs are estimated to be of the order $10^7-10^8 M_\odot$, as demonstrated by \citet{dasyra06a, dasyra06b} using VLT spectroscopic data and the M$_{BH} - \sigma$ relationship of \citet{msigma}.  The black hole masses derived from similar data on a dozen PG~QSOs agree with those of coalesced ULIRGs \citep{dasyra07}, suggesting that the bulk of the black hole growth takes place in the ULIRG phase of the merger.

Ultimately, if ULIRGs and PG~QSOs are linked through mergers, then we should see an evolution of the energy production mechanism along the merger sequence.  As the merger ages, the dominant source of infrared radiation should change from starburst (triggered by the interaction) to nuclear activity.  As part of QUEST, we performed \textit{Spitzer} IRS observations of PG~QSOs which showed that starbursts are responsible for at least $\sim$30\%, but likely most, of the far-infrared (FIR) luminosity of the quasars \citep{schweitz06}.   V09a found that, on average, AGN contribute $\sim 40$\% of the bolometric luminosity in the QUEST ULIRGs.  The AGN contribution ranges from $\sim$15\% -- 35\% among cool optically classified H~II-like and LINER ULIRGs, and $\sim$50 and $\sim$75\% among warm Seyfert 2 and Seyfert 1 ULIRGs, respectively. This number exceeds $\sim$80\% in PG QSOs.  Thus, a trend of increasing AGN contribution is seen along the merger sequence, in general agreement with the standard ULIRG-QSO evolution scenario. 

As ULIRGs are the results of the merging of gas-rich disk galaxies, obscuration may be an issue for optical and infrared observations.  
While radio emission is less affected by intervening material, the bolometric
luminosity in the radio band is insignificant to prove that accretion
onto supermassive black holes is the dominant energetic process.
X-ray emission, on the other hand, generally
contributes more significantly than the radio emission to the
bolometric luminosity of an AGN and is also less affected by obscuration than the optical and infrared emission. X-ray observations therefore have the
potential to isolate the AGN contribution from that of the
starburst, assuming the former arises from a compact region associated with the accretion disk. \textit{Chandra} and \textit{XMM-Newton} are excellent
complementary instruments for this purpose.  The high spatial
resolution of \textit{Chandra} is ideally suited to search for
unresolved hard X-ray (2--10 keV) nuclei indicative of AGNs.  However,
the absence of such a nucleus does not necessarily imply a starburst
origin since a large gas column ($N_H \geq 10^{24}$~cm$^{-2}$) in the line of sight (possibly due to a torus) could
strongly attenuate the hard X-rays.  In this case, X-rays emitted
along the polar axis may be electron scattered into the line of sight,
with the signature of an Fe~K$\alpha$ line of large equivalent width
\citep[$\sim$1~keV; e.g.,][]{ghm, krolik}.  This is where the
excellent sensitivity and spectral resolution of \textit{XMM-Newton}
near the Fe~K complex become most useful for AGN diagnostics.

In the present paper, we focus on the X-ray properties of the QUEST
ULIRGs and PG~QSOs.  There have been previous X-ray snapshot studies
of U/LIRGs \citep[e.g.,][]{frances, ptak03, teng}, but the sample
selection criteria for these surveys differed, making systematic
comparisons difficult.  The analyses of the X-ray data presented in this paper are performed in an uniform manner with an unprecedented large number of ULIRGs and PG~QSOs.  It is hoped that the contributions of the starburst and the AGN to the bolometric luminosity in these objects can be quantified systematically and with statistical significance.  Our current sample consists of 40 ULIRGs and
26 PG QSOs from the QUEST sample that have publicly available X-ray
data from {\it Chandra} and/or {\it XMM-Newton}.  These represent 51\% of the ULIRGs and 79\% of the PG~QSOs in the full QUEST sample.  Table~\ref{tab:sample}
lists the QUEST objects with X-ray data.  3C~273 and PG~0157+001 are both ULIRGs that are also PG~QSOs.  For the purpose of this paper, and to be consistent with the previous papers in this series (e.g., V09a), we consider these to be ULIRGs.  To improve the statistics of
our analysis and extend the range in infrared luminosity, we also add to
the sample 26 non-QUEST U/LIRGs from the \textit{Chandra} archive.
These objects are part of the Revised Bright Galaxy Survey
\citep[RBGS;][]{rbgs} and are discussed in more detail in 
Iwasawa et al. (2010, submitted).

The organization of this paper is as follows.  The techniques used for
taking and reducing the data are discussed in \S\ref{sec:obs}.  The
spectral fitting analyses for both ULIRGs and PG~QSOs are described in
\S\ref{sec:spectra}.  We address the origin of the soft excess seen in
PG~QSOs in \S\ref{sec:soft}.  In \S\ref{sec:multi}, we comment on the
possible evolutionary link between ULIRGs and quasars, combining the
X-ray data with recent \textit{Spitzer} and \textit{HST} observations.
The results are summarized in \S\ref{sec:summary}.  In Appendix A, we
examine the reliability of the hardness ratio (HR) method developed in
\citet{teng} for low-count X-ray sources.  Throughout this paper, we
adopt $H_0$ = 70 km s$^{-1}$ Mpc$^{-1}$, $\Omega_M$ = 0.3, and
$\Omega_\Lambda$ = 0.7.

\section{Observations and Data Reduction} 
\label{sec:obs}

The observations of ULIRGs and PG QSOs presented in this paper are selected from the {\it Chandra} and {\it XMM-Newton} archives as well as our own (PI: Veilleux) guest observer (GO) programs ({\it Chandra} cycle 10 and {\it XMM-Newton} cycle 7).  Only ACIS-S data are considered for the {\it Chandra} analysis.  Similarly, only EPIC data are considered for the {\it XMM-Newton} observations.  Table~\ref{tab:data} lists details on the available observations.  

\subsection{Data Calibration and Extraction}
\label{sec:extract}

\subsubsection{{\it Chandra Observations}}
\label{sec:cxoobs}

The reduction of the archived {\it Chandra} data was performed using CIAO version 4.1.1 and CALDB version 4.1.  The cycle 10 GO data were reduced using CIAO version 4.1.2 and CALDB version 4.1.3.  The Science Analysis Threads for ACIS data\footnote{http://cxc.harvard.edu/ciao4.1/index.html} outline the procedure used to process and reduce the data.  Table~\ref{tab:data} lists the total exposure for each observation after the selection of good time intervals where the data are not affected by background flares.  

The source spectra were extracted with circular regions centered on the source.  For most observations, background spectra were selected from an annular source-free region that surrounds the nuclear extraction area.  However, in particularly crowded fields, a nearby circular, source-free region was used.  The sizes of the regions vary depending on the angular extent of the sources, especially in U/LIRGs where there are a lot of extended diffuse emission.  The region sizes range between 15 to 50 arcseconds and were maximized to include all emission from the galaxies, unless limited by nearby sources or gaps in the CCD detectors.  

\subsubsection{{\it XMM-Newton Observations} }
\label{sec:xmmobs}

The archived {\it XMM-Newton} data were processed using the {\it XMM-Newton} Science Analysis System (SAS), version 7.1.0, released on 2007 July 8.  The event lists were re-calibrated with the latest available calibration files as of 2008 July.  The Cycle 7 GO data on PG~0838+770, PG~1435$-$067, and B2~2201+31A were reduced using SAS version 8.0.1. The standard processing procedures outlined in \S~4.13 of the {\it XMM-Newton} SAS User's Guide (Issue 5.0) were followed for both archived and GO data.  Times of high background flares were flagged and the total good time interval for each observation is listed in Table~\ref{tab:data}.  The standard method of background screening involves discarding time intervals affected by background flares where the background count rates at energies above 10~keV are above the recommended thresholds of 0.35 counts s$^{-1}$ and 1 counts s$^{-1}$ for EPIC-MOS and EPIC-pn data, respectively. While the MaxSNR method introduced by \citet{p04} would maximize the total net exposure times of the data, this method is only appropriate to use for data of high-flux sources.  X-ray observations of U/LIRGs \citep[e.g.,][]{frances, ptak03, teng} have shown that these data have low signal-to-noise ratios (SNRs).  Therefore, we have conservatively chosen to use the standard method of background screening for a consistent treatment of ULIRG and PG QSO data.  

The cross-calibration between the EPIC-MOS and the EPIC-pn cameras has some time and energy dependencies\footnote{{\it XMM-Newton} Calibration Documentation: http://xmm2.esac.esa.int/docs/documents/CAL-TN-0052-5-0.ps.gz.}.  Complications to the fits may result if spectra from both detectors are modeled simultaneously.  Therefore, we chose to model the spectra from only the EPIC-pn camera due to its high quantum efficiency and the inability to extract background spectra when the small window mode was used in many of the EPIC-MOS observations.  The EPIC-MOS data were used only in the few cases where the EPIC-pn data were unavailable or have much lower SNRs than the EPIC-MOS data:  PG~0838+770 has very low SNRs, PG~1244+026 has very few detected counts above $\sim$5~keV, PG~1613+658 and PG~1626+554 have data highly affected by background flares.

The SAS task EPATPLOT was used to determine whether the observations on the PG QSOs were affected by pile up.  Only three sets of observations were found to be piled-up: PG~0844+349, the 17 December 2004 observation of PG~1116+215, and PG~1426+015.  In these cases, the observations were re-extracted using annular regions to exclude the central part of the source which is the most susceptible to pile-up, as recommended by the SAS User's Guide.  This pile-up correction method is reliable and does not affect the shape of the output source spectrum\footnote{{\it XMM-Newton} Calibration Documentation: http://xmm2.esac.esa.int/docs/documents/CAL-TN-0036-1-0.ps.gz.}.  

The \textit{XMM-Newton} source spectra were also extracted with circular regions centered on the source ranging between 25 to 50 arcseconds depending on the angular sizes of the objects.  For EPIC-pn data, the pointing center falls near a CCD gap.  Thus, the background was selected from a nearby circular region in which no obvious background source resides.  The background regions for EPIC-MOS data were selected in a similar fashion.

Where there are multiple observations of the same source with the same instrument and filters, an average spectrum is created using the FTOOLS task MATHPHA giving each input spectrum equal weight.  The sources where this was performed are the EPIC-pn observations of PG~0050+124, PG~1116+215, PG~1440+356, and PG~1501+106.

\section{Spectral Analysis}
\label{sec:spectra}

The spectral analysis was performed using XSPEC v12.5.0.  All quoted errors are 90\% limits on one parameter ($\Delta \chi ^2$ or $\Delta$c-stat = 2.706).  The errors of the derived values in the rest of this paper are assumed to be at the 90\% confidence level.  Because of the differences in calibration, the effective energy range is 0.3--10.0 keV for EPIC-pn, 0.6-10.0~keV for EPIC-MOS, and 0.5--8.0~keV for ACIS.  For consistency between the different detectors and with the literature, the soft X-ray band measurements are made between 0.5--2.0~keV while the hard X-ray band measurements are made between 2.0--10.0~keV where the best-fit models are used to extrapolate the 0.5--0.6~keV and 8.0--10.0~keV measurements from the EPIC-MOS and ACIS observations, respectively.


\subsection{Results: PG~QSOs}
\label{sec:bin}
Depending on the number of detected counts, the extracted spectra of the PG~QSOs are binned differently for spectral modeling.  Their source spectra were binned to at least 50 counts per bin with the exception of those from fainter sources with relatively short integration times (PG~0838+770, PG~1001+054, PG~1004+130, PG~1126$-$041, PG~1244+026, PG~1309+355, PG~1411+442, PG~1426+015, PG~1435$-$067, PG~1613+658, PG~1626+554, and PG~2214+139) which were binned to at least 15 counts per bin so that $\chi ^2$ statistics would be applicable.  

In modeling the spectra, we take the same basic approach as in \citet{teng09}: first, we assume a simple power-law distribution absorbed only by the Galactic column to describe the emission from the AGN.  If the model is not a satisfactory fit to the data, then we consider adding a MEKAL component to describe the starburst, absorption by intervening material near the central source, and emission lines to model Fe~K$\alpha$ and lines of other elements, if applicable.  The F-test is used to determine whether the additional components to the basic model are significant\footnote{Under certain conditions, such as testing for a spectral line, it is inappropriate to use the F-test for model selection.  See \citet{ftest} for caveats and details.}.  We assume P$_{F-test} < 0.001$ for significant additions.  Hereafter, we refer to these models as the Power-Law (PL) models.  Table~\ref{tab:qsoscat} lists the best-fit parameters of the PL models to the PG~QSOs and these are shown with the spectra in Figure~\ref{fig:allspecs}.  More complex models involving reflection and scattering such as those presented in \citet{p05} (hereafter P05 models), the blurred ionized reflection model presented in \citet{crummy06} (C06), and a constrained version of the C06 model with fewer free parameters (C06con, described below) were also considered. 

The P05 model consists of a power law for the AGN plus four different continuum models to explain the soft excess often seen in AGN spectra (see \S~\ref{sec:soft}).  These include a blackbody, multi-color blackbody, bremsstrahlung, and a second power law.  Some of the sources also required additional absorption edge features between 0.6--0.8~keV.  The best-fitting model varies for each source and according to their study, there does not seem to be an universal model for the quasar spectra.  On the other hand, C06 selected a relativistically blurred reflection model as their universal model for the quasar continua.  The model is a relativistic convolution (KDBLUR in XSPEC) of a photoionized disk reflection model (REFLION).   The model assumes that a semi-infinite slab of optically thick cold gas of constant density is illuminated by emission from the corona.  The reflected and the direct components are then convolved with a Laor line profile \citep{laor} to create the effect of blurring from a relativistic accretion disk.  The blurred reflection model was invoked to explain the smooth soft excess features and the lack of strong iron lines seen in the PG~QSOs.  Unlike the P05 models, the C06 model is an universal model and can thus be applied to all the quasars in our sample.  We first checked that we have applied the C06 model properly by ensuring that the best-fit values are consistent with those published in C06 for observations that overlap both samples.  We then applied the C06 model more widely to all of our quasar spectra.  The range of best-fit parameters derived from the C06 models require extreme values for the accretion disk. For example, the disk emissivity index  ($\epsilon$, the power law dependence of the emissivity, $r^{-\epsilon}$)  ranges from 1.3 to 10.0, the upper limit of allowable values.  Therefore, we constrained the C06 model by fixing some of these parameters at their more widely accepted values to see if it would yield acceptable fits.  In our constrained version (C06con), the emissivity index of the disk in the KDBLUR model component is fixed at the more commonly accepted value of 3.0.  Since the iron abundance is less reliable, we conservatively fix the abundance of the REFLION component at solar.  For the majority of objects, the inclination of the disk is also fixed at the default value of 30 degrees in order for the fit to converge so that $\chi_\nu^2 < 2.0$.  We find that the C06 models perform significantly better than the C06con models in only a few instances.  

Tables~\ref{tab:qsop05}, \ref{tab:qsokdblurfree}, and \ref{tab:qsokdblur} list the best-fit parameters to the P05, C06, and C06con models to the PG~QSO data, respectively.  Figure~\ref{fig:chicompc}  shows comparisons of the reduced $\chi^2$ values between the PL and the P05 and C06 models; the PL models appear to be as good as, or better than, the P05 and C06 models in fitting the data.  In terms of the basic spectral properties, the PL models seem to be equally good in determining the photon index as both P05 and C06.  Figure~\ref{fig:gammacomp} is a comparison of the photon index as determined by the three methods.  The P05 and PL models appear to be consistent with each other; however, the PL model requires softer spectra than the C06 models.  Nevertheless, the photon indices as determined by all three methods are within the range previously measured in the PG~QSOs by other authors \citep[e.g.,][]{porquet04}. 

As the figures show, with the exception of a handful of observations, none of these models is statistically favored.  However, variability in quasar spectra is well known, so the ideal model of the data should also naturally explain the observed variability seen in these objects.  Thus, based on variability arguments, the C06 model appears to be the least likely explanation for the quasar spectra.  C06 evaluated only single-epoch quasar spectra from \textit{XMM-Newton}.  They chose the longest available observation for objects with multiple data sets in the archive.  
The PL models can easily explain the variability by a change in the absorbing column (see Table~\ref{tab:qsoscat}); physically, this can be due to a clumpy torus viewed approximately edge on \citep[see also][]{teng09}.  In order for the C06 model to explain the variability (particularly the significant variability of PG~0050+124 and PG~1501+106), large changes in $\Gamma$, the emissivity index, and the disk inclination are required (see Table~\ref{tab:qsokdblurfree}).  While it can be argued that the changes in $\Gamma$ and emissivity index are due to changes in the accretion flow of the black hole, the required change of about 30$^\circ$ in the observed inclination of the accretion disk over a the period of a few years is hard to justify.

\subsection{Results: ULIRGs}
\label{sec:ulirgsspec}

As with previous U/LIRG surveys \citep[e.g.,][]{frances, ptak03,
  teng}, most ULIRGs in the present study were detected, but with many
fewer counts than the average PG~QSO.  We separated them into three
``brightness'' categories: weak, moderate, and strong for the purpose
of the analysis.

The ``weak'' sources are those with count rates $\lesssim 0.01$ and 0.05 counts per second when observed by \textit{Chandra} and \textit{XMM-Newton}, respectively.  With the time allocated, these sources do not have enough counts for traditional spectral fitting and their spectral properties and fluxes are measured using the HR method \citep[see Appendix A and][for details]{teng}.  The spectral properties derived using the HR method for the weak ULIRGs are listed in Table~\ref{tab:hr}.  

The ``moderate'' sources are those with relatively low count rates, but the exposure times are long enough to obtain low signal-to-noise spectra with more than 100, but less than 1000, counts.  For these sources, their spectral properties are modeled using both the HR method and traditional spectral fitting.  The traditional spectral fits were performed using the Cash Statistics \citep[c-stat][]{cash} option in XSPEC on unbinned spectra.  The details of the fitting procedure are the same as those in \citet{teng08}.  Only the PL models are applied to these low signal-to-noise spectra.  The best-fit properties are presented in Table~\ref{tab:ulirgfits}. 

Finally, the ``strong'' sources are those with relatively high count rates and relatively high signal-to-noise spectra having more than 1000 counts.  The spectral modeling was performed using the $\chi^2$ statistics option in XSPEC as the spectra were binned to at least 15 counts per bin, with the exception of 3C~273, which was binned to at least 100 counts per bin.  Again, only the PL models were applied to these ULIRG spectra.  The spectral properties of these sources are also listed in Table~\ref{tab:ulirgfits}.

In general, the 2--10~keV band is less likely to be affected by
obscuration and this is where AGN emission, if present, would dominate.
Figure~\ref{fig:lircomp} is a comparison of the 2--10~keV and infrared
luminosities of the U/LIRGs and PG~QSOs.  The U/LIRGs include the 26
RBGS objects from the \textit{Chandra} archive for a total of 66
objects.  For AGN-dominated objects like the PG~QSOs, the 2--10~keV
luminosity is consistently about 3\% of the infrared luminosity.  The
2--10~keV luminosity of the U/LIRGs, however, vary between 0.0001 to
3\% of the infrared luminosity with the majority of the ULIRGs falling
between $\log$($L_{2-10 keV}/L_{IR}$) of (--4.5, --1.5).  
Comparison with the effective mid-infrared optical depth,
$\tau_{eff}$, derived by V09a suggest that AGN-dominated objects
also have the smallest extinction. 

We also compared the 2--10~keV luminosity with the bolometric
luminosity of the U/LIRGs and PG~QSOs; the 2--10~keV to bolometric
luminosity ratio is our proxy for AGN dominance.  The bolometric
luminosity is defined as $L_{bol} = 1.15 L_{IR}$ for the U/LIRGs and
$L_{bol} = 7 L_{5100 \AA} + L_{IR}$ for the PG~QSOs.  The left panel
in Figure~\ref{fig:lbolcomp} plots this ratio against the total
bolometric luminosity of our sample.  For the U/LIRGs there is simply
a small shift in both axes from Figure~\ref{fig:lircomp}.  The
right-hand panel shows the same ratio plotted against the AGN
contribution to the bolometric luminosity as derived by V09a.  Again,
we observe a spread of three orders of magnitude in the 2--10~keV to
bolometric luminosity for the U/LIRGs.  The importance of nuclear
activity relative to that of the starburst increases slightly with the
bolometric (infrared) luminosity of the ULIRG. We find again that the
less obscured objects are more likely to be AGN-dominated.

\citet{vei09b} derived black hole mass estimates using the {\it H}-band elliptical host magnitude to black hole mass relation from \citet{marconi}.  The derivation does not include dust extinction beyond the nuclear regions of the hosts; this may possibly lead to underestimates of the black hole masses.  The presence of recent/ongoing non-nuclear star formation is also not excluded which may cause overestimates in the black hole masses.  
Applying these photometrically derived black hole mass estimates from \citet{vei09b}
to our X-ray observations, we calculate the 2--10~keV-to-Eddington
luminosity ratio, our proxy for the Eddington ratio.
Figure~\ref{fig:eddcomp} compares the Eddington ratio derived from the
X-ray methods with that from the mid-infrared methods presented in
V09a.  Since the X-ray values are only the 2--10~keV luminosity and
not the bolometric luminosity of the AGN, there is a shift of a factor
of about 30 -- 100 between the horizontal and vertical values.  The
two methods are linearly correlated for the PG~QSOs, but a discrepancy
between the X-ray and mid-infrared AGN diagnostic methods is observed
for the U/LIRGs.  The cause of this discrepancy is discussed in
\S~\ref{sec:methods}.

\section{An Universal Spectral Model for the PG~QSOs}
\label{sec:soft}

PG~QSOs have been very well studied in X-rays and at other
wavelengths.  However, many X-ray studies on PG~QSOs in the literature
have focused on only a handful of objects.  Most studies have found
that the X-ray spectra of PG~QSOs tend to be more or less featureless,
with sometimes small emission features near the iron K complex arising
from neutral or ionized iron.  Often, these iron emission lines have
relatively small equivalent widths ($\sim$100~eV) and are generally
narrow.  The continua of the X-ray spectra can be fit by a power law
with $\Gamma \sim$1.8 for the AGN, but the X-ray spectra of several
PG~QSOs (as well as many Seyfert galaxies) also contain what is termed
the ``soft excess'' where there is emission in excess of the power-law
below $\sim$2~keV.  \citet{caixa} uses a model-independent method of
comparing the 0.5--2~keV and 2--10~keV flux ratios as an independent
measure of the soft excess strength.  The authors confirm that the
soft excess is a common feature in low-redshift active galaxies. 

The presence of the soft excess in active galaxies was first identified in the 1980s \citep[e.g.,][]{arnaud}.  Ever since, the origin of the emission has remained a puzzle.  In recent years, several authors including \citet{porquet04}, P05, C06 have performed systematic analyses on large groups of quasar spectra in an attempt to find the best model that describes the excess emission.  The current leading models for the soft excess include Compton up-scattering \citep{porquet04}, blurred reflection model (C06), complex absorption \citep{gd04, sobo07}, and other models such as discussed in P05.  

\citet{porquet04} at first thought that the soft excess originates from the inner accretion disk.  This model explains naturally the smooth transition from UV accretion disk emission to the soft excess.  However, the inferred blackbody temperatures from the modeling are too hot to be explained by direct emission from a thin accretion disk for reasonable values of the black hole mass.  Thus, they prefer their alternate model, the Compton up-scattering of the extreme ultraviolet photons from the accretion disk to form the soft excess.

C06 presented a blurred reflection model as a universal model for the PG~QSO spectra.  Blurred reflection, caused by the relativistic motion in the accretion disk, is invoked due to the lack of broad iron lines observed in these objects.  While this model fits the spectra well, it also requires extreme values for some of the model parameters, as explained in \S~\ref{sec:bin}.

\citet{gd04} suggested that the soft excess is caused by a broad absorption trough at $\sim$2--5~keV, possibly related to an accretion disk wind.  The smoothness of the observed soft excess, however, cannot be reproduced in the latest simulations \citep[e.g.,][]{dp09}.  \citet{sobo07} favor a complex absorption model based on their spectral modeling of two AGNs (PG~1211+143 and 1H0707--495) with large observed soft excesses.

Instead of a universal model for the soft excess, \citet{p05} found in their survey of 42 PG~QSOs that the X-ray continua of these quasars are well fit by a combination of four different models.  The four models are blackbody, multicolor blackbody, bremsstrahlung, and power law.

From a statistical stand point, all of the above models (as well as the PL models presented in \S~\ref{sec:bin}) are equivalent.  The fitting statistics of the models to the data are very similar.  Below, we describe a new approach to help us discriminate between the various models for the soft excess, where we model all of the PG~QSOs with a single universal model, varying only a few key parameters from object to object.  We do not know the actual origin of the soft excess, so instead of making assumptions about which model describes physics that we do not fully understand, we look at the data in a more general way.  

The soft excess is a common feature in low-redshift active galaxies, including quasars.  Therefore, it can be assumed that all QSO spectra have the same basic shape arising from the same physical phenomena.  Let us further assume that there is a single (yet unknown) ideal model that describes the quasar spectra in the form of a power law (for the AGN) plus an additional component.  Thus, each observed quasar spectrum is then a random variation of the basic ideal model.  If we fit all the spectra simultaneously with the same model, then we should be able to define a median model for the quasars as a class.  This median model, then, should help us identify a favored model that best fits {\em all} objects.

\subsection{Modeling PG~QSOs as a Class}
\label{sec:class}

For this multi-source fitting, we chose to use only
\textit{XMM-Newton} EPIC-pn observations to minimize cross-calibration
issues.  This includes 22 objects except PG~0804+761 (undetected),
PG~1244+026 (few counts above $\sim$5~keV), PG~1613+658, and
PG1626+554 (no pn data).  Average spectra for objects with multiple
observations were created using the FTOOLS task MATHPHA and used for
the multi-source fitting.

For most of the models we tested using this global fitting method, we treated the continuum model as having two components and all of these components are modified by Galactic absorption.  Since most of the PG~QSO spectra do not show strong iron lines (see Figure~\ref{fig:allspecs})\footnote{The lack of strong iron lines in PG~QSOs may be due to the X-ray Baldwin Effect where the intensity of the line is inversely correlated with the total luminosity of the quasars \citep{baldwin} or gravitational smearing (C06).}, we only applied the global model to the continuum.  The first of these components is AGN emission which includes the standard power law for an AGN spectrum.  The second is the soft-excess component.  This can be a blackbody, a Comptonization model, or a reflection model.  For all the objects, we linked the photon index, soft-excess temperature, and the normalizations for each component to be the same for all objects ({\it i.e.} the median model for the class).  However, we allowed model parameters that describe characteristics that may vary in individual sources, such as disk inclination, ionization parameter, and intrinsic source absorption, to be free.  Since each source also has different brightnesses, a multiplicative factor was included for each of the components to adjust for the differences in intensity amongst the sources as well as the relative contributions between the different model components.  Simply, the global model can be described in equation form:
\begin{math}
\rm Model_{\it i} = Abs_{Gal, \it i} \times Abs_{int, \it i} \times [X_{\it i} \times AGN + Y_{\it i} \times SE ],
\end{math}
where X$_i$ is the multiplicative factor for the AGN component that varies depending on the source $i$ and Y$_i$ is a similar factor for the soft excess component.

\subsection{The Favored Model}
\label{sec:prefmod}

Table~\ref{tab:modstat} summarizes the 13 spectral models we examined as possible universal models for the PG~QSOs.  We first tested a simple power law model for the global QSO fit, model A of Table~\ref{tab:modstat}.  It is a very poor fit to the data, giving a reduced $\chi^2 \sim$8.0.  We then tested the absorption-based models for the origin of the soft excess, models B--E.  A scattering model (B), see \citet{teng09} for a description, provided a better fit ($\chi^2_\nu \sim$3.0) but the model offered a poor description of the data at higher energies.  A single partial covering absorption model (C) also gave a very poor fit ($\chi^2_\nu \sim$3.4).  An additional partial covering absorber (D) provided a much better fit ($\chi^2_\nu \sim$1.8); however, the model severely underestimated the spectral flux above $\sim$5~keV.  In this model, the nominal photon index of the spectrum was $\sim$2.8, much steeper than the generally observed range for AGNs (1.6--2.2).  By adding a MEKAL component for the soft excess to the double-partial-covering model (E), we derived an even better model to the data ($\chi^2_\nu \sim$1.5), but the model still underestimated the flux above $\sim$5~keV as shown in Figure~\ref{fig:allpndel}.  

We then considered the Comptonization model of \citet[model F;][]{porquet04} for the soft excess.  This model gives $\chi^2_\nu \sim$2.1.  While this is a much better fit to the data than the simple power law models, large residuals remain, providing a poor fit both at low ($\lesssim$0.5~keV) and at high ($\gtrsim$4~keV) energies.  

Finally, we tested the reflection-based models for the soft excess, models G--M.  The reflection-based models are XSPEC models PEXRAV (a neutral reflector), PEXRIV (an ionized reflector), and a blurred reflection model (the C06 model).  REFLION component in the C06 model is more complex than PEXRIV.  PEXRIV considers only bound-free transitions in the reflected spectrum; REFLION also includes the ionization states and transitions for O and Fe ions.  We first modeled each reflection-based model (models G--I).  These attempts produced very poor results ($\chi^2_\nu \gtrsim$3.0) and suggest that reflection cannot explain the soft excess.  We then added a redshifted blackbody component to each of the reflection models for the soft excess.  The reduced $\chi^2$ value for the model J is $\sim$1.3, a great improvement over the previous models.  This model describes well the spectra above $\sim$1~keV, but is a poor model below this energy (see Figure~\ref{fig:allpndel}).  The PEXRIV plus blackbody model (model K) appeared to be a good fit to the data ($\chi^2_\nu \sim$1.2; Figure~\ref{fig:allpndel}), except for an absorption feature around 0.7~keV (the atomic transition of O~VII or O~VIII).  Lastly, we evaluated the C06 model plus a blackbody (model L).  Surprisingly, this more complex model for an ionized reflector gives a worse fit than the simpler model K ($\chi^2_\nu \sim$1.4; Figure~\ref{fig:allpndel}).  In particular, this model is a very poor description of the spectra above $\sim$4~keV. 

Based on the statistics and the residuals of the models, the multi-source fitting method indicates that a reflection-based model is favored as the universal model.  However, there is still a requirement for a blackbody component in order to model the soft excess.  One reason why the PEXRIV component offers a better fit than the REFLION component may be because many of the spectra do not require ionized reflection.  Only 10/22 objects (45\%) require the ionization parameter to be above 30~erg~cm~s$^{-1}$, the minimum value for this parameter in REFLION.  The PEXRIV component allows a full range of ionization parameters, starting with a minimum value of 0~erg~cm~s$^{-1}$.  It should be noted that model K with the PEXRIV component requires a steeper photon index (2.37) than model L with the REFLION component (2.02).  While a photon index of $\sim$2.4 is steeper than the generally accepted value, it is the median value seen in PG~QSOs (see Figure~\ref{fig:gammacomp}).  When a redshifted absorption edge centered at around $\sim$0.68~keV was added to model K, the statistics greatly improved ($\chi^2_\nu \sim$1.1; Figure~\ref{fig:pexrivdel}).  For the three objects with EPIC-MOS data that were not included in the multi-source fitting, we applied the best-fit global model with the AGN and soft excess components fixed to individually model the MOS spectra.  The best global fit of model M implies that $\Gamma = 2.37 \pm 0.02$ and $kT = 0.127 \pm 0.001$~keV with an absorption edge at $0.68\pm 0.01$~keV and optical depth of $0.31^{+0.02}_{-0.03}$.  The nominal energy of the absorption edge is consistent with the atomic transitions of O~VII and O~VIII.  The reflected portion of the AGN contribution is $\sim$6--37\%, with a median value of 18\%, of the power law luminosity for all 25 objects\footnote{A reflection percentage of 37 is rather high.  The two sources that have 37\% reflected fraction are PG~1001+054 and PG~1004+130.  The former object is the quasar with the fewest detected counts by EPIC-pn.  Its lack of counts, particularly in the hard band where the reflected component dominates, may have resulted in a poor fit.  The PG~1004+130 spectrum used in the universal fit is the average of two epochs.  As can be seen in Figure~\ref{fig:allspecs} and unlike other objects presented in this paper with multiple observations, there is noticeable change in the spectral shape, particularly at below 2~keV.  The average spectrum may not be representative of the source spectrum in general.  Also, \citet{miller} studied this object extensively.  There is no obvious soft excess in this object.}.  Some of the other parameters derived from model M for the PG~QSOs are listed in Table~\ref{tab:pexrivstat}.

\subsubsection{Disk Inclination}
\label{sec:discinc}

C06 suggested that the \textit{XMM-Newton} data are sufficient to robustly measure the inclination angle of the accretion disk.  
Neither of the distributions of inclination angles from the global PEXRIV nor the constrained C06 models from the same 22 EPIC-pn spectra matches a random distribution.  The inclination measurements are highly dependent on the model, which may indicate that the data cannot adequately assess the inclination values.  This may also be a result of the small number of quasars in the sample.  In any case, we caution against relying too heavily on the inclination measurements derived from the models whether fitted to individual or multiple sources. 
Since the subsequent analysis is based mainly on good X-ray flux measurements, we choose the model M as the best-fit model despite its shortcomings. 

\subsubsection{Origin of the Soft Excess}
\label{sec:sesource}

The global modeling of the quasar spectra suggests that the soft excess
is not well described by an absorption model (\S~\ref{sec:prefmod}).
Figure~\ref{fig:ktvplref} clearly shows that the 0.5--10~keV flux from the soft excess component is correlated with that from the AGN component based on model M.  Linear regression analysis
suggests that the two components are linearly correlated: $\log L_{\rm
  bbody} = (1.04 \pm 0.28) \log L_{\rm pl} - (2.69 \pm 12.28)$.  The
correlation is very strong with $R^2 = 0.65$ and a significance of $>
99.99$\%.  The outlier, PG~0050+124 (I~Zw~1), is removed from the regression
analysis; recall that this is the only source that retains an
absorption feature in the residuals after the inclusion of the 0.7~keV
absorption edge (see Figure~\ref{fig:pexrivdel}).  To within the errors, the expression derived from the PG~QSO data is consistent with that found in a sample of \textit{Swift}/BAT AGNs with soft excess \citep[$R^2 = 0.48$;][]{winter09}, but with a greater correlation coefficient.  PG~1501+106 is the only overlap between the BAT and our PG~QSO samples.  The soft excess in AGNs and PG~QSOs seems to
arise from the same process.  The linear relationship between the
blackbody and power law luminosities precludes absorption as the
origin of the soft excess, in agreement with \citet{winter09}.
A starburst origin of the soft excess in these objects is also ruled out.  The soft excess luminosities are much higher than the expected starburst X-ray luminosities based on FIR measurements of star-forming galaxies \citep{persic04}.  

Since the global model fits the AGN and the soft excess components
independent of each other, it does not require the AGN components to
be correlated with the soft excess.  Therefore, the linear
relationship between the input power law and the soft excess
luminosities is significant and implies a link between the source of
the power-law emission and the soft
excess. 
We also compared the soft excess luminosity with black hole mass and ionization state of the reflector.  Neither of these quantities is apparently correlated with the soft excess.

Many authors \citep[e.g.,][]{gd04, winter09} have argued that if the
soft excess were due to thermal emission arising from the accretion
disk, then the blackbody temperature should correlate with the mass of
the black hole or the Eddington ratio.  Indeed, in this scenario, the
thermal temperature should scale according to $T \propto M^{-1/4}
(L/L_{Edd})^{1/4}$, assuming that all of the gravitational energy gained from accretion is re-radiated by the disk.  Contrary to this, previous studies and our analysis in
\S~\ref{sec:bin} have found that the thermal temperature of the soft
excess is consistently $\sim$0.1~keV.  The constancy of the
observed temperature is unexpected since the masses of both Seyferts
\citep{winter09} and PG~QSOs \citep{vei09b} span two orders of
magnitude ($\sim 10^7 - 10^9$M$_\odot$).  Using the absorption
corrected 2--10~keV to Eddington luminosity ratio as a proxy for the
Eddington ratio, we find that this quantity also spans two orders of
magnitude for both PG~QSOs and Seyferts
\citep[Figure~\ref{fig:eddcomp} and][]{winter09}.
Figure~\ref{fig:temphist} is a histogram of the predicted PG~QSO disk
temperatures calculated using the photometric black hole mass
estimates of \citet{vei09b} and the 2--10~keV luminosities.  These are
only estimates because the hard X-ray to bolometric luminosity
correction for AGNs is uncertain (e.g., extreme ultraviolet measurements are not available for most of these objects).  \citet{lusso} found in 150 COSMOS AGNs that the 2--10~keV to bolometric luminosity correction can range from about 10 to about 200 with some dependence on the Eddington ratio.  For our sources, given the weak dependence of the temperature on the Eddington ratio, the predicted  disk temperatures may be underestimated by a factor of a few at most.   As the figure shows, the disk temperature is
surprisingly constant for all objects, with a mean of
$\sim$1~eV and a standard deviation of $\sim$0.4~eV.  Only a fraction
of all QSOs could be plotted in this figure due to the lack of black hole mass information for many objects in our sample \citep[see][]{vei09b}, so small number
statistics may be an issue here.  Nevertheless, the predicted
temperature of the disk seems too low to explain the measured soft
excess temperature of $\sim$0.1~keV.

The mechanism producing the soft excess must be mass independent. For example, suppose the corona is geometrically thick and thus at a temperature comparable to the local virial temperature in all sources.  Then $T\propto\sqrt{M/R}$, which at a given number of gravitational radii or a given orbital speed (e.g., the speed characteristic of the broad line region) is the same for any mass, hence Compton up scattering could produce a characteristic spectrum that is independent of mass.  Therefore, coronal effects may contribute to the soft excess.  However, the main argument against this scenario is that the virial temperatures corresponding to broad line region velocities (a few $\times 10^3$~km~s$^{-1}$) are much too high (a few keV) to produce the soft excess.  

The multi-source fitting has demonstrated that all the models discussed have difficulties explaining all the properties of the soft excess emission. 

\section{The Multi-wavelength Properties of the QUEST ULIRGs and PG QSOs}
\label{sec:multi}

\subsection{FIR Classes of PG~QSOs}

\citet{netzer} separated PG~QSOs into two different FIR classes, strong and weak FIR emitters, based on the 60-to-15 $\mu$m luminosity ratio.  The authors found that both types of QSOs have similar underlying AGN spectral energy distributions (SEDs) in the infrared ($>$ 1 $\mu$m).  We find that this statement seems also valid in the X-rays: The appears to be no correlation with the FIR strength with the AGN contribution to the infrared luminosity (Figure~\ref{fig:lircomp}). The C06 models suggest that the strong FIR-emitters have softer spectra (due to the possible contribution from a starburst to the 0.5--2~keV band) than the weak FIR-emitters, but these models are not favored for the reasons discussed in \S~\ref{sec:prefmod}.

\subsection{AGN Contribution to the Bolometric Luminosity}
\label{sec:methods}

V09a presented six independent mid-infrared methods for
determining the AGN contribution to the overall bolometric luminosity
of ULIRGs and QSOs as part of the \textit{Spitzer}-QUEST study.  The
details of the methods are presented in the Appendix of V09a
and not repeated here. Since the hard X-ray emission is dominated by
the AGN (assuming no Compton-thick sources), we use the ratio of the absorption-corrected 2--10~keV
luminosity to the bolometric luminosity as a proxy for the AGN
contribution to the bolometric luminosity and compare the results with
those derived by V09a. The results on the QSOs and ULIRGs
are presented in Figure~\ref{fig:agnpervedd}.

Figure~\ref{fig:agnpervedd} shows a positive correlation between the \textit{Spitzer}-derived AGN contribution and the absorption-corrected 2--10 keV to bolometric luminosity ratio, although considerable scatter is evident in the relation. Note also that the range of hard X-ray to bolometric luminosity ratios is considerably broader than the range of \textit{Spitzer}-derived AGN contributions. The \textit{Spitzer} results are likely more uncertain in objects with intermediate and high $\tau_{eff}$ (as discussed in V09a), but Figure~\ref{fig:agnpervedd} shows no link between $\tau_{eff}$ and the discrepancy between the \textit{Spitzer} and X-ray results. We discuss two possible causes for this discrepancy, aside from the obvious and likely possibility of real intrinsic variations in the hard X-ray to bolometric luminosity ratio among pure AGN \citep[see Figure~11 of][]{lusso, just}.

\subsubsection{Zero-point Calibrations}

One potential source of uncertainty is the definitions of the
zero-points for the mid-infrared methods in V09a.  The pure-AGN zero
points were derived from the FIR-undetected PG~QSOs.  Following
\citet{netzer}, the lack of FIR detection was assumed to indicate a
lack of starburst in these objects.  As discussed in \citet{netzer},
this assumption is likely accurate to $\pm$10-30\% and thus cannot
account for the substantially larger discrepancies between the X-ray
and mid-infrared methods.  The pure-starburst zero points were derived
from HII ULIRGs, which are known to be different in terms of gas
density and radiation fields from less luminous, optically-selected
starbursts.  For instance, the $f_{30}/f_{15}$~micron flux ratios and
the 7.7~micron PAH equivalent widths derived by \citet{brandl} for a
sample of optically-selected starburst nuclei are lower than those
found in the HII ULIRGs.  If we were to use the \citet{brandl} values
as the pure-starburst zero points, then the estimated AGN
contributions from the mid-infrared data would be {\em systematically}
reduced, but this still could not explain the very broad range in hard
X-ray to bolometric luminosity ratio observed among ULIRGs and QSOs.

\subsubsection{A Matter of Obscuration}

A more likely explanation for the discrepancy between the two methods
is related to the fact that the hard X-ray and mid-infrared
observations probe very different regions of the AGN.  The hard X-rays
represent direct or reflected emission from the accretion disk of the
black hole or from material near it.  The X-ray source occupies a
small volume, on the scale of less than a parsec.  In contrast, the
mid-infrared diagnostic method of measuring AGN emission relies on the
detection of high-ionization fine-structure line or PAH emission which
is produced from a larger volume (tens to hundreds of parsecs).  Thus, the mid-infrared emitting region is less likely
to be affected by obscuration than the central X-ray emitting
region. The broad range in hard X-ray to bolometric luminosity ratio
may be due to unsuspected obscuration that is not correlated with the
mid-infrared extinction measured on a larger scale
(Figures~\ref{fig:lircomp} and \ref{fig:agnpervedd}).  Both \textit{Chandra} and \textit{XMM-Newton} operate at
0.5--10~keV, below the peak of the Compton reflection hump, the
detection of which can better constrain the absorbing column.  Mrk~273
is a good case in point.  \citet{teng09} had a marginal detection of
Mrk~273 above 10~keV with \textit{Suzaku}.  The simultaneous modeling
of the \textit{Suzaku}, \textit{Chandra}, and \textit{XMM-Newton}
spectra found that the source is highly obscured.  The absorption
corrected 2--10~keV luminosity with the \textit{Suzaku} data is $\sim
3.2$ times that derived from \textit{Chandra} or \textit{XMM-Newton}
data alone.  Mrk~273 is also one of the sources that exhibits moderate
extinction from the mid-infrared data ($\tau_{eff} \sim 6.4$).  This
is a change of about half an order of magnitude or $\sim 30$\% in AGN
percentage (Figure~\ref{fig:lircomp}). Mrk~231 is another, more
extreme, example: The detection of hard X-rays above 10 keV by
\citet{braito} boosted the absorption-corrected hard X-ray luminosity
by more than an order of magnitude, bringing it to a value consistent
with that of powerful QSOs.
In summary, the absorption-corrected hard X-ray luminosities of ULIRGs
and, to a lesser extent, QSOs are uncertain, so it is not
inconceivable that the very broad range of hard X-ray to bolometric
luminosity ratios of ULIRGs and QSOs is due at least in part to
unsuspected obscuration of the hard X-ray source.

\subsection{Trends with Merger Phase}
\label{sec:evolve}  

While the hard X-ray to bolometric luminosity ratio may not be an
accurate {\em absolute} measure of the AGN contribution to the
bolometric luminosity of ULIRGs and QSOs, trends of this ratio within
these classes of objects may provide useful {\em relative}
information.

In the evolutionary scenario of \citet{sanders88}, starburst-dominated
``cool'' U/LIRGs evolve into AGN-dominated ``warm'' ULIRGs and then
eventually optically selected quasars.  In this scenario, the AGNs
turn on only near the end of the merging process.  The recent {\em
Spitzer} results of V09a have brought some support to this
scenario. The X-ray data also seem to be largely consistent with this
picture.  The top panel of Figure~\ref{fig:propcomp} supports previous
arguments that ``warm'' (``cool'') objects like Seyfert 1 ULIRGs and
PG QSOs (HII and LINER ULIRGs) are AGN-dominated
(starburst-dominated).  ULIRGs with intermediate infrared colors also
have intermediate hard X-ray to bolometric luminosity ratios.  This
supports the idea that the cool, starburst-dominated, and obscured
objects evolve to become warm, AGN-dominated, ``naked'' quasars.

The middle panels of Figure~\ref{fig:propcomp} demonstrate that the
more X-ray luminous AGN are more likely to be sources optically
classified as either a QSO or a Seyfert galaxy.  The HII ULIRGs have
the lowest hard X-ray to bolometric luminosity ratios on average, as
expected.  The LINER ULIRGs show the broadest range of 2--10~keV to
bolometric luminosity ratios, suggesting that the energy source of
infrared-selected LINERs is a wide mixture of AGN and starburst
\citep{sturm, vei09a}.  The very low $L_{2-10 keV}/L_{bol}$ in some
ULIRGs may be a sign of unsuspected large X-ray absorptions.

\citet{vei02} defined interaction classes (IC) for merging systems where class I are systems on their first approach.  The galaxies are unperturbed, with no evidence of tidal features.  The IC II represents sources that are in the first-contact phase of the interaction where strong bars or tidal tails have yet to be formed.  The pre-merger stage, or IC III, consists of systems that show identifiable binary nuclei and strong tidal features such as tails and bridges.  This class is further divided into two sub-categories: IC IIIa are sources that are wide binaries where the nuclei are $>$ 10~kpc (projected) apart and IC IIIb are sources that are close binaries where the nuclei are $\leq$ 10~kpc (projected) apart.  IC IV represents the merger stage of the interaction.  These sources have prominent tidal tails.  The sub-category of IVa (diffuse mergers) includes sources that have diffuse central regions whereas IVb (compact mergers) includes sources that are dominated by a single nucleus.  The division between IVa and IVb represents the difference between binary and single nucleus objects.  Finally, IC V is synonymous with old mergers.  These sources do not have strong tidal features, but their central morphologies are disturbed, similar to those of objects in IC IV.  Therefore, the sequence of IC I--V represents the complete merging sequence of galaxy systems.  It is of note that almost all 1-Jy ULIRGs are classified as IC III--V (moderate age mergers) and the PG QSOs as IC IVb--V (old mergers).  

The bottom panels of Figure~\ref{fig:propcomp} suggest that AGN activity is
often most dominant in coalesced remnants corresponding to the latest stage
of the merger -- IVb and V.  The steep increase of AGN dominance
between classes IIIa/b (early stages of the merger) and IV/V suggests
that black hole growth picks up in the post-merger phase of the
interaction, as found in the {\em Spitzer} data.  The scatter among
each interaction class indicates that stochastic events may trigger
nuclear activity at any time along the merger sequence. However, the
likelihood of finding AGN-dominated U/LIRGs increases along the merger
sequence and in objects with warmer infrared colors ($f_{25}/f_{60}$).

It is interesting to compare these results with the predictions of
recent numerical simulations \citep[e.g.,][and references
therein]{hopkins08}. These models suggest that the starburst dominates
the total luminosity prior to and during the merger \citep[phase D
of][]{hopkins08}.  After coalescence, the central black hole in these
simulations grows rapidly before the ``blowout'' phase (phase E) where
an AGN-driven wind is purported to expel the remaining dust and gas,
removing material for both accretion and star formation.  The result
is a luminous, blue quasar with little star formation (phase F).  The
accretion rate of the active nucleus is predicted to peak between
phases D and E. Then the luminosity of the quasar fades during the
post-blowout quasar stage.  For comparison,
Figure~\ref{fig:comp_model} shows the evolution of the AGN luminosity
along the final stages of the merger sequence derived from the
mid-infrared and X-ray data.  The data indicate that star formation
peaks prior to the coalescence of the nuclei, while the accretion rate
onto the black hole increases rapidly to peak at coalescence.  There
is no significant change in the quasar luminosity after coalescence.
The fading of the quasar must happen after the epoch covered by the
QUEST sample of QSOs. The purported galactic-scale winds in these
simulations are observed in several QUEST ULIRGs \citep[e.g.,
][and references therein]{rupke05a, rupke05b, vei05}.

\section{Summary}
\label{sec:summary}

We have performed an uniform analysis of X-ray data on 40 ULIRGs and
26 PG~QSOs from the QUEST sample.  New and archival observations
obtained using \textit{Chandra} and \textit{XMM-Newton} were used for
this study. The X-ray results were compared with those recently
derived from {\em Spitzer} IRS spectra. The major conclusions are as
follows:

\begin{enumerate}

\item By fitting the PG~QSO spectra simultaneously, we favor a
  reflection-based model with $\Gamma$ = 2.37 $\pm$ 0.02 and a
  black-body component with $kT = 0.127 \pm 0.001$ keV to account for
  the ubiquitous soft excess.  The best-fit universal model indicates
  that the soft excess luminosity/flux is linearly related to the
  0.5--10~keV absorption-corrected power-law luminosity/flux.  This
  implies that the source of the soft excess is directly related to
  accretion onto the central black hole, rather than an external
  factor such as intervening absorption.  An absorption
  edge at 0.68 $\pm$ 0.01 keV with optical depth of
  0.31$^{+0.02}_{-0.03}$ is required for the universal spectral model to fit the QSO spectra as a class. This edge
  is consistent with the atomic transitions of O~VII and O~VIII.

\item There does not appear to be any correlation between the FIR
  emission strength of the QSOs and the 2--10~keV to bolometric luminosity ratio.  This extends to the X-rays the conclusion of
  \citet{netzer} that there is no obvious difference in the underlying
  infrared AGN SED of strong and weak FIR emitting PG~QSOs.

\item Using the absorption-corrected hard X-ray luminosity as a proxy
  for the AGN contribution to the bolometric luminosity in these
  systems, we find results that generally agree qualitatively with
  those from V09a.  The ratio of absorption-corrected hard X-ray to
  bolometric luminosity is not an accurate {\em absolute} measure of
  the AGN contribution to the bolometric luminosity of ULIRGs and
  QSOs, but trends of this ratio within these classes of objects
  provide useful {\em relative} information.  The likelihood of
  powerful nuclear activity increases along the merger sequence and in
  objects with warmer dust temperatures and AGN optical
  signatures. The scatter in these trends is likely due to stochastic
  accretion events, unsuspected X-ray absorption, or variations in the
  intrinsic hard X-ray to bolometric luminosity ratio of pure AGNs.

\item The bolometric luminosity of the AGN in U/LIRGs and PG~QSOs
  evolve with merger stage. The starburst seems to dominate the total
  power prior to the merger. Then the accretion rate onto the black
  hole increases rapidly during coalescence at which point the AGN
  dominates the bolometric luminosity.  The predictions from numerical
  simulations are largely consistent with these results. 
\end{enumerate}

\acknowledgements

We thank the anonymous referee for his or her careful reading of the manuscript and for providing useful comments that improved the paper.  Thanks are due to Chris Reynolds, Lisa Winter, and Dave Hunter for
useful discussions.  We are also grateful to Richard Mushotzky, Cole
Miller, and Andy Ptak who provided invaluable suggestions on an
earlier draft of this paper.  The hardness ratio script was written by
Andy Young.  Last but not least, we thank Andrew Wilson posthumously
for providing the inspiration for this project. This research is based
on observations obtained with \textit{XMM-Newton}, an ESA science
mission with instruments and contributions directly funded by ESA
Member States and NASA.  We made use of the NASA/IPAC Extragalactic
Databased (NED), which is operated by the Jet Propulsion Laboratory,
Caltech, under contract with NASA.  We acknowledge support by NASA
through \textit{Chandra} General Observer grant GO90125X,
\textit{XMM-Newton} General Observer grant NNX09AF05G, and the
Astrophysics Data Analysis Program (ADP) research grant NNX09AC79G,
all to the University of Maryland.


\clearpage

\begin{center}
  {\bf APPENDIX A} --- The Reliability of the Hardness Ratio Method
\end{center}

As discussed in \S~\ref{sec:ulirgsspec}, U/LIRGs are notoriously difficult to observe in the X-ray.  Most of these objects are faint, either due to the lack of an AGN or the presence of heavy obscuration.  
Traditional spectral fitting cannot be used to model the complex spectra of these sources when counts are limited.  The hardness ratio method developed in \citet{teng} has proven to be effective in finding obscured AGNs in at least one case.  \textit{IRAS}~F04103--2838 was found to contain both a starburst and an AGN with only 30 detected counts in a 10~ksec \textit{Chandra} observation.  A deeper ($\sim$20~ksec) \textit{XMM-Newton} follow-up revealed an Fe~K$\alpha$ line at rest-frame energy of $\sim$6.5~keV, consistent with cold neutral iron \citep{teng08}.  The hard X-ray emission is dominated by a nearly Compton-thick AGN with intrinsic 0.2--10~keV luminosity $\sim$10$^{43-44}$~ergs~s$^{-1}$. 

Since the detected counts of the majority of objects in the U/LIRG surveys are low, the errors associated with the HR method are inherently large.  To further test the reliability of the HR method in recovering the input spectrum, we have performed a set of simulations.  In these simulations, we set the input model of an unabsorbed AGN at $z \sim$0.1 as a redshifted power law with $\Gamma$ at 1.8 (the canonical value for AGNs) absorbed by a Galactic column of $2 \times 10^{20}$~cm$^{-2}$.  The normalization of the input power law model was such that the model 0.5--10~keV flux is $\sim 5 \times 10^{-14}$~ergs~s$^{-1}$~cm$^{-2}$, the mean value observed in faint U/LIRGs.  

We first tested the dependence of the HR method on the number of detected counts by varying the exposure times in the simulations.  For each exposure time tested, the average of 1000 simulations showed that the nominal $\Gamma$ and the 0.5--2~keV and 2--10~keV flux values are remarkably stable even when the ``detected'' counts were reduced to as low as 30 (a 5~ksec exposure).  On average, the output $\Gamma$ determined from the HR method remained the same as the input, but the error bars increased as the number of detected counts decreased.  The nominal 0.5--2 and 2--10~keV fluxes are within 1\% of those from the input model.  The top portion of Table~\ref{tab:hrsims} summarizes these results.

In the HR method, we assumed that the only absorption is from the Galaxy, but this is not the case for the U/LIRGs we observe.  The internal absorption in these objects are often high.  Therefore, we also tested the dependence of the HR method on the intrinsic absorption of the source.  To this end, we added intrinsic source absorption to the input spectrum.  For 15~ksec exposures, we varied the intrinsic source absorption between 10$^{20-22}$~cm$^{-2}$ by steps of $5 \times 10^{20}$~cm$^{-2}$ to see how the measured spectral parameters using the HR method are affected.  Figure~\ref{fig:hrsims} shows the results of these simulations.  As the internal column increases, the photon index becomes flatter and then inverted ($\Gamma < 1$).  $\Gamma$ begins to become inverted when the column is $\gtrsim 5 \times 10^{21}$~cm$^{-2}$ as seen in the bottom portion of Table~\ref{tab:hrsims}.  Since the 0.5--2~keV flux is more readily affected by absorption, it deviates from the input spectrum by more than 10\% when the absorption is at only $1 \times 10^{21}$~cm$^{-2}$.  On the other hand, the 2--10~keV flux is more stable, but begins to deviate from the input spectrum when the absorption is high enough that $\Gamma$ becomes inverted.  Of course, if the source has a softer spectrum (e.g., containing a commonly seen soft-excess component), then it would require a higher column for the spectrum to become inverted.

One of the important usages of the HR method is to estimate the photon index of the X-ray spectrum which in turn gives estimates of the 0.5--2 and 2--10~keV fluxes of our targets.  The spectral index is the parameter that is the basis for the flux estimates.  As shown by the simulations, the accuracy of $\Gamma$ based on the HR method depends on the intrinsic absorption of the source.  Figure~\ref{fig:hrgamma} compares the $\Gamma$ derived from the HR method and the traditional method for the moderately bright U/LIRGs in \S~\ref{sec:ulirgsspec} and in the archived \textit{Chandra} sample discussed in Appendix~B.  The figure shows that our measurements from the two methods are consistent with each other unless the intrinsic column densities are $\gtrsim 10^{22}$~cm$^{-2}$, in agreement with the simulations.  The next consideration is to see how well the estimated fluxes from the HR method match those of the traditional method.  We have plotted in Figure~\ref{fig:hrcomp} the 0.5--2~keV and 2--10~keV fluxes derived from both the HR and the traditional fitting method with more complex spectral models for the moderately bright U/LIRGs in \S~\ref{sec:ulirgsspec}.  As the figure shows, the majority of the HR flux values from a simple unabsorbed power law model are within 50\% of the spectral fitting values of more complex models.  The HR method is more likely to overestimate the fluxes in both the soft and hard bands.  The medians values for $F_{HR}/F_{fits}$ are approximately 1.2 and 1.3 for the soft and hard bands, respectively.

As \citet{teng09} and other authors have demonstrated, the X-ray spectra of U/LIRGs are often more complex than a simple power law model can characterize.  We further tested the HR method by assuming that the input model contains a reflection component or that the intrinsic absorption is due to a partial covering absorber.  The addition of a reflection or a partial covering component flattens the AGN power law spectrum, mimicking the effects of high intrinsic absorption.  For an intrinsic absorption of $1 \times 10^{22}$~cm$^{-2}$, the HR method recovers the input flux very well for both the reflection and partial covering models\footnote{The reflection model assumes a power law plus a PEXRAV component.  The normalization of the reflected component is set to 2\% of the intrinsic power law component as measured in the complex \textit{Suzaku} spectrum of {\it IRAS}~F05189--2524 \citep{teng09}.  The photon index of the power law spectrum is fixed at 1.8 and the recovered $\Gamma$ is 0.70$^{+0.32}_{-0.30}$.  For the partial covering model, the covering factor is assumed to be 90\%, similar to that found in Mrk~273 \citep{teng09}.  In this case, the input $\Gamma$ is again fixed at 1.8 and the HR estimate of $\Gamma$ is 0.91$^{+0.30}_{-0.28}$.}.  In fact, for both models, the recovered 0.5--2~keV fluxes are within 2\% of the input model flux.  On the other hand, the HR estimates of the 2--10~keV fluxes are above the input values by 56.4 and 35.3\% for the reflection and partial covering models, respectively.  These errors are comparable to or better than the estimates obtained for the simple power law model at the same column density (Table~\ref{tab:hrsims}).  Therefore, even for objects with complex spectra, the HR method is able to provide fair approximations of their spectral properties.

\begin{deluxetable}{clcccrcl}
\tablecolumns{8}
\tabletypesize{\scriptsize}
\tablecaption{The Sample}
\tablewidth{0pt}
\tablehead{
\colhead{Galaxy} & \colhead{$z$} & \colhead{log($\frac{L_{bol}}{L_\sun}$)} & \colhead{Type} & \colhead{IC} & \colhead{NS} &\colhead{N$_{\rm H,~Gal}$}& \colhead{Ref} \\
\colhead{(1)} & \colhead{(2)} & \colhead{(3)} & \colhead{(4)} & \colhead{(5)} & \colhead{(6)} & \colhead{(7)} & \colhead{(8)}
}
\startdata
\cutinhead{ULIRGs}
                 F00091$-$0738  & 0.118  & 12.36  &  HII  &   IIIb  &     2.31  &      3.23&1 \\
                 F00188$-$0856  & 0.128  & 12.43  &    L  &      V  &$<$  0.34  &      3.21&2  \\
              F00456$-$2904:SW  & 0.110  & 12.29  &  HII  &   IIIa  &    22.80  &      1.68&1  \\
                 F01004$-$2237  & 0.118  & 12.36  &  HII  &      V  &$<$  0.32  &      1.58&2 \\
              F01166$-$0844:SE  & 0.118  & 12.15  &  HII  &   IIIb  &    10.78  &      4.67&1 \\
                  PG0157$+$001  & 0.163  & 12.69  &   S1  &    IVb  &$<$  0.42  &      2.56&2, 3, 4 \\
                 F02021$-$2103  & 0.116  & 12.13  & none  &    IVa  &$<$  0.32  &      1.55&1\\
              IRAS03521$+$0028  & 0.152  & 12.62  &    L  &   IIIb  &     3.86  &      12.5&2 \\
\enddata
\tablerefs{1 = {\it Chandra} AO 10 object; 2 = Teng et al. 2005; 3 = Piconcelli et al. 2005; 4 = Crummy et al. 2006; 5 = Teng et al. 2008; 6 = Ptak et al. 2003; 7 = Imanishi 2004; 8 = Teng et al. 2009; 9 = Franceschini et al. 2003; 10 = Sanders et al. (2003); 11 = {\it XMM-Newton} AO 7 object}
\tablecomments{
Col.(1): Galaxy name.  Coordinate-based names beginning with "F" are sources in the IRAS Faint Source Catalog.  Col.(2): Redshift.  Col.(3): Bolometric luminosity.  For ULIRGs, we assume $L(bol) = 1.15 L(IR)$.  For PG QSOs, we assume $L(bol) = 7 L(5100~\AA) + L(IR)$ \citep[see][for details on $L(5100~\AA)$]{netzer}.  Col.(4): Optical spectral type, from Veilleux et al. (1995, 1999a) and Rupke et al. (2005a).  Col.(5): Interaction class, from Veilleux et al. 2002, Veilleux et al. 2006, or Veilleux et al. 2009 (in order of preference).  Col.(6): Nuclear separation, in kpc.  Col.(7): Galactic column density from Dickey and Lockman (1990) in units of 10$^{20}$ cm$^{-2}$.  Col.(8): References for X-ray data.
}
\label{tab:sample}
\end{deluxetable}

\begin{deluxetable}{clcllr}
\tablecolumns{6}
\tabletypesize{\scriptsize}
\tablecaption{Journal of Observations}
\tablewidth{0pt}
\tablehead{
\colhead{Galaxy} & \colhead{Date} & \colhead{Telescope} & \colhead{Obs ID} & \colhead{PI} & \colhead{Exposure} \\
\colhead{(1)} & \colhead{(2)} & \colhead{(3)} & \colhead{(4)} & \colhead{(5)} & \colhead{(6)}}
\startdata
\cutinhead{ULIRGs}
                 F00091$-$0738  & 2008 November 1  & CXO  &  70188510342  &   Veilleux &     15.2\\
                 F00188$-$0856  & 2003 September 4  & CXO  &    7007814102 &   Wilson  & 9.8 \\
                 & 2004 December 20 & XMM  &  0200630201 &   Imanishi & 38.0 \\
              F00456$-$2904:SW  & 2009 May 22 & CXO &  70188610343 &   Veilleux&    15.1 \\
                 F01004$-$2237  & 2003 August 3  & CXO &  7007824103&      Wilson  & 9.4\\
              F01166$-$0844:SE  & 2008 October 31& CXO&  70188710344 &   Veilleux&    15.8\\
                  PG0157$+$001  & 2000 July 29  & XMM  &   0101640201&  Aschenbach & 4.5  \\
                  & 2003 August 23  & CXO  &   7007834104&   Wilson &10.6 \\
\enddata

\tablecomments{
Col.(1): Galaxy name.  Coordinate-based names beginning with "F" are sources in the IRAS Faint Source Catalog.  Col.(2): Observation start date.  Col.(3): Telescope used (CXO = {\it Chandra}; XMM = {\it XMM-Newton}).  Col.(4): Observation identification number.  Col.(5): Principal investigator of the observation.  Col.(6): Net exposure time in ks after selection of good time interval.
}

\tablenotetext{a}{3C~273 is often observed by {\it XMM-Newton} for calibration purposes.  We selected the six observations that have the longest integration times and also maximize the period of elapsed time between the first and final observations.}
\label{tab:data}
\end{deluxetable}

\begin{deluxetable}{clccccccccccc}
\tablecolumns{13}
\setlength{\tabcolsep}{0.03in}
\tabletypesize{\tiny}
\tablecaption{PG~QSOs: PL Model Best-fit Values}
\tablewidth{0pt}
\tablehead{
\colhead{Galaxy} & \colhead{Obs. Date} & \colhead{$kT$} & \colhead{$\Gamma$} & \colhead{$N_H$} & \colhead{E$_{\rm line}$} & \colhead{$\sigma_{\rm line}$} &\colhead{EW} &\colhead{$\chi^2_\nu$ (d.o.f.)} &\colhead{$F_{0.5-2}$} &\colhead{$F_{2-10}$} &\colhead{$L_{0.5-2}$} & \colhead{$L_{2-10}$}\\
\colhead{(1)} & \colhead{(2)} & \colhead{(3)} & \colhead{(4)} & \colhead{(5)} & \colhead{(6)} & \colhead{(7)} &\colhead{(8)} &\colhead{(9)} &\colhead{(10)} & \colhead{(11)} &\colhead{(12)} & \colhead{(13)}}
\startdata
                  PG0050$+$124  & 2002 Jun 22\tablenotemark{a}& \nodata  &  2.31$^{+0.02}_{-0.02}$ & 0.04$^{+0.01}_{-0.03}$&\nodata  & \nodata&\nodata&1.22 (630)&8.38$^{+9.31}_{-0.07}$&8.25$^{+7.57}_{-0.07}$&0.96& 0.74\\
		  & 2005 Jul 18\tablenotemark{a}&\nodata&2.34$^{+0.009}_{-0.007}$&56.86$^{+18.15}_{-14.59}$&\nodata&\nodata&\nodata&1.09 (812)&5.16$^{+0.03}_{-0.03}$&4.93$^{+0.11}_{-0.17}$&0.93&0.68\\
                  PG0838$+$770  & 2009 Mar 2 & \nodata&1.49$^{+0.08}_{-0.08}$&\nodata&\nodata&\nodata&\nodata&1.06 (138)&0.29$^{+0.03}_{-0.04}$&0.78$^{+0.08}_{-0.08}$&0.14&0.35\\
                  PG0844$+$349  & 2000 Nov 5 & 0.15$^{+0.05}_{-0.06}$& 2.66$^{+0.05}_{-0.06}$  & 6.13$^{+3.03}_{-1.39}$&6.4(f)& 0.0(f)& 0.11$^{+0.19}_{-0.11}$&0.86 (264)&6.55$^{+0.13}_{-0.21}$&5.13$^{+0.31}_{-0.36}$&1.42&0.63\\
		  & & &   & &6.7(f)& 0.0(f)& 0.18$^{+0.21}_{-0.17}$&&&&&\\
                  PG0953$+$414  & 2001 Nov 22  & 0.16$^{+0.03}_{-0.07}$ & 2.44$^{+0.03}_{-0.03}$  &18.52$^{+9.84}_{-5.66}$  & \nodata  &\nodata&\nodata&1.04 (324) & 3.39$^{+0.05}_{-0.07}$&2.93$^{+0.17}_{-0.19}$&10.89&6.51\\
                  PG1001$+$054  & 2003 May 4& 0.09$^{+0.01}_{-0.01}$& 2.01$^{+0.67}_{-0.48}$&8.09$^{+5.47}_{-3.57}$&\nodata & \nodata&\nodata&0.96 (19)&0.02$^{+0.01}_{-0.01}$&0.12$^{+0.05}_{-0.09}$&0.10&0.12\\
                  PG1004$+$130  & 2003 May 4  &\nodata  &  1.67$^{+0.20}_{-0.11}$&2.99$^{+2.67}_{-1.37}$  &\nodata  & \nodata  & \nodata & 1.04 (111) & 0.09$^{+0.01}_{-0.01}$&0.31$^{+0.06}_{-0.08}$&0.30&0.57\\
\enddata
\tablenotetext{a}{The spectrum requires additional components for redshifted absorption edges originating from O~VII or O~VIII.}  

\tablecomments{
(f) denotes a fixed value.  Col.(1): Galaxy name.  Col.(2): Observation start date.  Col.(3): Gas temperature from the MEKAL component in units of keV.  Col.(4): Photon index of the direct and indirect power-law components.  Col.(5): Hydrogen column density within the source in units of $10^{22}$~cm$^{-2}$.  Col.(6): Rest-frame energy of the iron emission line(s) in units of keV.  Col.(7): Width of the iron line(s) in units of keV.  Col.(8): The equivalent width of the iron line(s) in units of keV.  Col.(9): Reduced $\chi^2$ of the best-fit model followed by the number of degrees of freedom in ().  Col.(10): Observed 0.5--2~keV flux in units of $10^{-12}$~erg~s$^{-1}$~cm$^{-2}$.  Col.(11): Observed 2--10~keV flux in units of $10^{-12}$~erg~s$^{-1}$~cm$^{-2}$.  Col.(12): Nominal absorption corrected 0.5--2~keV luminosity in units of $10^{44}$~erg~s$^{-1}$.  Col.(13): Nominal absorption corrected 2--10~keV luminosity in units of $10^{44}$~erg~s$^{-1}$.
}
\label{tab:qsoscat}
\end{deluxetable}

\begin{deluxetable}{clccccccccccc}
\tablecolumns{13}
\setlength{\tabcolsep}{0.01in}
\tabletypesize{\tiny}
\rotate
\tablecaption{PG~QSOs: P05 Model\tablenotemark{a} Best-fit Values}
\tablewidth{0pt}
\tablehead{
\colhead{Galaxy} & \colhead{Obs. Date} & \colhead{$kT$} & \colhead{$\Gamma$} & \colhead{$N_H$} & \colhead{E$_{\rm line}$} & \colhead{$\sigma_{\rm line}$} &\colhead{EW} &\colhead{$\chi^2_\nu$ (d.o.f.)} &\colhead{$F_{0.5-2}$} &\colhead{$F_{2-10}$} &\colhead{$L_{0.5-2}$} & \colhead{$L_{2-10}$}\\
\colhead{(1)} & \colhead{(2)} & \colhead{(3)} & \colhead{(4)} & \colhead{(5)} & \colhead{(6)} & \colhead{(7)} &\colhead{(8)} &\colhead{(9)} &\colhead{(10)} & \colhead{(11)} &\colhead{(12)} & \colhead{(13)}}
\startdata
                  PG0050$+$124  & 2002 Jun 22\tablenotemark{b}& 0.07$^{+0.007}_{-0.007}$; 0.21$^{+0.007}_{-0.011}$  &  2.25$^{+0.05}_{-0.03}$ & 0.09$^{+0.08}_{-0.02}$&6.4(f)  & 0.0(f)&0.04$^{+0.06}_{-0.04}$&1.13 (626)&8.38$^{+1.07}_{-1.52}$&8.33$^{+0.11}_{-0.63}$&1.09& 0.76\\
		  & 2005 Jul 18&0.08$^{+0.002}_{-0.002}$; 0.25$^{+0.01}_{-0.01}$&2.09$^{+0.03}_{-0.03}$&0.04$^{+0.03}_{-0.02}$&6.4(f)&0.0(f)&0.02$^{+0.03}_{-0.02}$&1.10 (812)&5.16$^{+0.20}_{-1.24}$&4.89$^{+0.15}_{-0.44}$&0.60&0.44\\
                  PG0844$+$349  & 2000 Nov 5 & 0.08$^{+0.01}_{-0.01}$; 0.19$^{+0.03}_{-0.03}$& 2.06$^{+0.07}_{-0.09}$  & \nodata&\nodata&\nodata&\nodata&0.84 (266)&6.51$^{+0.32}_{-0.57}$&5.37$^{+0.31}_{-0.39}$&0.73&0.53\\
                  PG0953$+$414  & 2001 Nov 22\tablenotemark{c} & 2.11$^{+0.32}_{-0.14}$ & 2.57$^{+0.02}_{-0.02}$  &\nodata  & \nodata  &\nodata&\nodata&1.06 (326) & 3.40$^{+0.07}_{-0.06}$&2.98$^{+0.19}_{-0.20}$&5.81&4.87\\
                  PG1001$+$054  & 2003 May 4& \nodata& 2.06$^{+0.15}_{-0.16}$&8.05$^{+4.34}_{-1.48}$\tablenotemark{d}&\nodata & \nodata&\nodata&1.79 (21)&0.02$^{+0.004}_{-0.007}$&0.10$^{+0.03}_{-0.04}$&0.08&0.09\\
\enddata
\end{deluxetable}

\begin{deluxetable}{clcccccccccccc}
\tablecolumns{14}
\tabletypesize{\tiny}
\rotate
\tablecaption{PG~QSOs: C06 Model Best-fit Values}
\tablewidth{0pt}
\tablehead{
\colhead{Galaxy} & \colhead{Obs. Date} & \colhead{$\Gamma$} & \colhead{$N_H$} &\colhead{$\epsilon$}& \colhead{$R_{\rm in}$} & \colhead{$i$} &\colhead{Fe}&\colhead{$\xi$} &\colhead{$\chi^2_\nu$ (d.o.f.)} &\colhead{$F_{0.5-2}$} &\colhead{$F_{2-10}$} &\colhead{$L_{0.5-2}$} & \colhead{$L_{2-10}$}\\
\colhead{(1)} & \colhead{(2)} & \colhead{(3)} & \colhead{(4)} & \colhead{(5)} & \colhead{(6)} & \colhead{(7)} &\colhead{(8)} &\colhead{(9)} &\colhead{(10)} & \colhead{(11)} &\colhead{(12)}&\colhead{(13)}&\colhead{(14)}}
\startdata
                  PG0050$+$124  & 2002 Jun 22\tablenotemark{a}&  2.27$^{+0.01}_{-0.01}$ & 0.04$^{+0.005}_{-0.005}$&10.0$^{...}_{-2.0}$&3.5$^{+0.3}_{-0.4}$ & 31.0$^{+2.8}_{-3.3}$&0.1$^{+0.1}_{...}$&2132$^{+1544}_{-881}$&1.20 (625)&8.35$^{+7.65}_{-3.68}$&8.26$^{+7.00}_{-3.50}$&0.83& 0.74\\
		  & 2005 Jul 18&2.23$^{+0.01}_{-0.01}$&0.05$^{+0.003}_{-0.003}$&5.1$^{+0.3}_{-0.2}$&2.7$^{+0.2}_{-0.2}$&3.5$^{+7.8}_{-3.4}$&0.1$^{+0.1}_{...}$&1012$^{+61}_{-284}$&1.15 (812)&5.16$^{+44.84}_{-0.50}$&4.79$^{+38.26}_{-0.26}$&0.49&0.43\\
                  PG0838$+$770  & 2009 Mar 2 & 1.36$^{+0.04}_{-0.04}$&\nodata&5.0$^{+2.5}_{-1.6}$&1.2$^{+2.5}_{...}$&50.6$^{+9.1}_{-13.9}$&0.1$^{+0.3}_{...}$&300$^{+14}_{-157}$&1.08 (141)&0.33$^{+1.89}_{-0.24}$&0.75$^{+0.46}_{-0.51}$&0.15&0.34\\
                  PG0844$+$349  & 2000 Nov 5 & 2.18$^{+0.04}_{-0.04}$  & \nodata&7.1$^{+1.0}_{-0.9}$&1.3$^{+0.3}_{-0.1}$&60.2$^{+3.0}_{-3.1}$&0.5$^{+0.3}_{-0.1}$& 521$^{+172}_{-85}$&0.85 (264)&6.55$^{+3.75}_{-4.92}$&5.23$^{+2.54}_{-3.75}$&0.65&0.52\\
                  PG0953$+$414  & 2001 Nov 22  & 2.27$^{+0.02}_{-0.02}$  &\nodata  &6.1$^{+0.9}_{-0.8}$& 1.4$^{+0.2}_{-0.2}$ &59.5$^{+3.4}_{-3.0}$&0.1$^{+0.1}_{...}$&81$^{+24}_{-15}$&1.02 (322) & 3.40$^{+2.60}_{-0.88}$&3.01$^{+1.51}_{-0.91}$&5.56&4.92\\
                  PG1001$+$054  & 2003 May 4& 1.01$^{+0.13}_{-0.01}$&\nodata&10.0$^{...}_{-4.3}$&3.0$^{+0.3}_{-1.1}$ & 22.2$^{+3.6}_{-4.1}$&6.7$^{+3.3}_{-2.8}$&305$^{+9.8}_{-2.2}$&1.41 (17)&0.03$^{+0.04}_{-0.02}$&0.12$^{+0.40}_{-0.85}$&0.02&0.09\\
\enddata

\tablenotetext{a}{The spectrum requires additional components for redshifted absorption edges originating from O~VII or O~VIII.}

\tablenotetext{b}{The inclusion of a Gaussian for the emission line at 6.4~keV is required.}

\tablecomments{
(f) denotes a fixed value.  Col.(1): Galaxy name.  Col.(2): Observation start date.  Col.(3): Photon index of the direct and indirect power-law components.  Col.(4): Hydrogen column density within the source in units of $10^{22}$~cm$^{-2}$.  Col.(5): The power law dependence of the emissivity ($r ^{-\epsilon}$).  Col. (6): Inner radius of the accretion disk in gravitational radii with the outer radius fixed at the default value (100 gravitational radii).  Col.(7): Inclination of the accretion disk to the line of sight in degrees.  Col. (8): Iron abundance of the reflector relative to the Solar value.  Col.(9): Ionization parameter of the gas.  Col.(10): Reduced $\chi^2$ of the best-fit model followed by the number of degrees of freedom in ().  Col.(11): Observed 0.5--2~keV flux in units of $10^{-12}$~erg~s$^{-1}$~cm$^{-2}$.  Col.(12): Observed 2--10~keV flux in units of $10^{-12}$~erg~s$^{-1}$~cm$^{-2}$.  Col.(13): Nominal absorption corrected 0.5--2~keV luminosity in units of $10^{44}$~erg~s$^{-1}$.  Col.(14): Nominal absorption corrected 2--10~keV luminosity in units of $10^{44}$~erg~s$^{-1}$.
}
\label{tab:qsokdblurfree}
\end{deluxetable}

\begin{deluxetable}{clcccccccccc}
\tablecolumns{12}
\setlength{\tabcolsep}{0.03in}
\tabletypesize{\tiny}
\tablecaption{PG~QSOs: Constrained C06 Model\tablenotemark{a} Best-fit Values}
\tablewidth{0pt}
\tablehead{
\colhead{Galaxy} & \colhead{Obs. Date} & \colhead{$\Gamma$} & \colhead{$N_H$} & \colhead{$R_{\rm in}$} & \colhead{$i$} &\colhead{$\xi$} &\colhead{$\chi^2_\nu$ (d.o.f.)} &\colhead{$F_{0.5-2}$} &\colhead{$F_{2-10}$} &\colhead{$L_{0.5-2}$} & \colhead{$L_{2-10}$}\\
\colhead{(1)} & \colhead{(2)} & \colhead{(3)} & \colhead{(4)} & \colhead{(5)} & \colhead{(6)} & \colhead{(7)} &\colhead{(8)} &\colhead{(9)} &\colhead{(10)} & \colhead{(11)} &\colhead{(12)}}
\startdata
                  PG0050$+$124  & 2002 Jun 22\tablenotemark{b}&  2.26$^{+0.02}_{-0.02}$ & 0.04$^{+0.002}_{-0.003}$&4.5(f)  & 30.0(f)&10000(f)&1.16 (633)&8.37$^{+0.34}_{-0.23}$&8.27$^{+0.24}_{-0.13}$&0.98& 0.75\\
		  & 2005 Jul 18\tablenotemark{b}&2.16$^{+0.01}_{-0.01}$&\nodata&1.9$^{+0.3}_{-0.2}$&87.7$^{+1.8}_{-0.6}$&2870$^{+720}_{-801}$&1.05 (810)&5.16$^{+0.61}_{-2.34}$&4.87$^{+0.56}_{-1.91}$&0.55&0.44\\
                  PG0838$+$770  & 2009 Mar 2 & 1.49$^{+0.09}_{-0.09}$&\nodata&4.5(f)&30.0(f)&30(f)&1.06 (137)&0.29$^{+0.05}_{-0.03}$&0.77$^{+0.17}_{-0.05}$&0.14&0.35\\
                  PG0844$+$349  & 2000 Nov 5 & 2.32$^{+0.04}_{-0.04}$  & \nodata&1.7$^{+0.6}_{-0.5}$&30.0(f)& 999$^{+45}_{-387}$&1.15 (267)&6.76$^{+2.55}_{-3.32}$&4.58$^{+2.23}_{-2.34}$&0.76&0.45\\
                  PG0953$+$414  & 2001 Nov 22  & 2.30$^{+0.02}_{-0.02}$  &\nodata  & 1.9$^{+0.9}_{-0.7}$ &30.0(f)&48$^{+16}_{-12}$&1.04 (324) & 3.39$^{+1.02}_{-0.71}$&2.93$^{+0.82}_{-0.75}$&5.79&4.79\\
                  PG1001$+$054  & 2003 May 4& 2.0(f)&\nodata&1.2 & 30.0(f)&297&4.80 (21)&\nodata&\nodata&\nodata&\nodata\\
                  PG1004$+$130  & 2003 May 4  &  1.40$^{+0.05}_{-0.05}$&\nodata  &4.5(f)  & 30.0(f)  & 30$^{+299}_{...}$ & 1.08 (111) & 0.10$^{+0.04}_{-0.02}$&0.33$^{+0.34}_{-0.09}$&0.18&0.57\\
		  &2005 Jan 5 & 1.24$^{+0.09}_{-0.10}$&0.14$^{+0.08}_{-0.07}$&4.5(f)&30.0(f)&30(f)&1.10 (96)&0.09$^{+0.01}_{-0.02}$&0.43$^{+0.04}_{-0.06}$&0.19&0.74\\
\enddata
\tablenotetext{a}{The accretion disk emissivity index of the kdblur component and the iron abundance of the reflion component are fixed at 3.0 and Solar, respectively.  The error estimates of individual parameters, fluxes, and luminosities for poor fits (i.e. $\chi^2_\nu > 2.0$) are not calculated.}
\tablenotetext{b}{The spectrum requires additional components for redshifted absorption edges originating from O~VII or O~VIII.}

\tablenotetext{c}{The inclusion of a Gaussian for the emission line at 6.4~keV is required.}

\tablecomments{
(f) denotes a fixed value.  Col.(1): Galaxy name.  Col.(2): Observation start date.  Col.(3): Photon index of the direct and indirect power-law components.  Col.(4): Hydrogen column density within the source in units of $10^{22}$~cm$^{-2}$.  Col.(5): Inner radius of the accretion disk in gravitational radii with the outer radius fixed at the default value (100 gravitational radii).  Col.(6): Inclination of the accretion disk to the line of sight in degrees.  Col.(7): Ionization parameter of the gas.  Col.(8): Reduced $\chi^2$ of the best-fit model followed by the number of degrees of freedom in ().  Col.(9): Observed 0.5--2~keV flux in units of $10^{-12}$~erg~s$^{-1}$~cm$^{-2}$.  Col.(10): Observed 2--10~keV flux in units of $10^{-12}$~erg~s$^{-1}$~cm$^{-2}$.  Col.(11): Nominal absorption corrected 0.5--2~keV luminosity in units of $10^{44}$~erg~s$^{-1}$.  Col.(12): Nominal absorption corrected 2--10~keV luminosity in units of $10^{44}$~erg~s$^{-1}$.
}
\label{tab:qsokdblur}
\end{deluxetable}

\begin{deluxetable}{clcrrrccccccc}
\tablecolumns{13}
\setlength{\tabcolsep}{0.03in}
\tabletypesize{\tiny}
\tablecaption{Hardness Ratio Estimates of Weak$^\dag$ ULIRGs}
\tablewidth{0pt}
\tablehead{
\colhead{Galaxy} & \colhead{Date} & \colhead{Extended?} & \colhead{Total} & \colhead{Hard} & \colhead{Soft} & \colhead{HR} &\colhead{$\Gamma$} &\colhead{N$_{H}$}&\colhead{$F_{0.5-2~keV}$}&\colhead{$F_{2-10~keV}$}&\colhead{$L_{0.5-2~keV}$}&\colhead{$L_{2-10~keV}$}\\
\colhead{(1)} & \colhead{(2)} & \colhead{(3)} & \colhead{(4)} & \colhead{(5)} & \colhead{(6)} & \colhead{(7)}& \colhead{(8)} & \colhead{(9)} & \colhead{(10)} & \colhead{(11)} & \colhead{(12)}&\colhead{(13)}}
\startdata
                 F00091$-$0738  & 2008 Nov 1  & No  &  7 &   3$^{+3.6}_{-1.6}$&     4$^{+3.2}_{-1.9}$& --0.14$^{+0.59}_{-0.33}$&1.03$^{+0.66}_{-1.13}$&0.44$^{+1.47}_{-0.40}$&0.11&0.56&0.04&0.20\\
                 F00188$-$0856  & 2003 Sep 4  & No  &16 & 6$^{+3.6}_{-2.4}$  & 10$^{+4.3}_{-3.1}$ &--0.25$^{+0.36}_{-0.25}$&1.1$^{+0.5}_{-0.6}$&0.36$^{+0.64}_{-0.31}$&0.42&1.69&0.20&0.71\\
                 & 2004 Dec 20 & \nodata  &  164 &   52$^{+8.8}_{-7.7}$ & 112$^{+10.7}_{-9.7}$ &--0.37$^{+0.09}_{-0.08}$&1.45$^{+0.16}_{-0.17}$&0.14$^{+0.09}_{-0.02}$&0.42&1.12&0.20&0.47\\
              F00456$-$2904: SW  & 2009 May 22 & Yes &  43&   9$^{+4.1}_{-2.9}$&    34$^{+6.9}_{-5.8}$&--0.21$^{+0.22}_{-0.17}$&1.15$^{+0.31}_{-0.41}$&0.35$^{+0.35}_{-0.21}$&0.74&3.11&0.22&0.94 \\
                 F01004$-$2237  & 2003 Aug 3  & No &  20&  6$^{+3.6}_{-2.4}$  & 14$^{+4.8}_{-3.7}$&--0.40$^{+0.32}_{-0.24}$&1.4$^{+0.6}_{-0.6}$&0.16$^{+0.47}_{...}$&0.26&3.64&0.09&1.29\\
                 F02021$-$2103  & 2009 Oct 12& No&  39   &18$^{+5.3}_{-4.2}$&      21$^{+5.7}_{-4.6}$&--0.08$^{+0.20}_{-0.16}$&0.89$^{+0.29}_{-0.36}$&0.55$^{+0.36}_{-0.23}$&0.56&3.56&0.19&1.21\\
              IRAS03521$+$0028  & 2002 Dec 25  & No&  3& 1$^{+2.3}_{-0.8}$ &   2$^{+2.6}_{-1.3}$ & --0.33$^{+1.23}_{-0.54}$&1.5$^{+1.9}_{-3.4}$&0.25$^{+5.64}_{...}$&0.14&0.33&0.09&0.20\\
\enddata

\tablecomments{$^\dag$ We arbitrarily consider a source ``weak'' when its count rate is $\lesssim$ 0.01 and 0.05 counts per second when observed with \textit{Chandra} and \textit{XMM-Newton}, respectively.
Col.(1): Galaxy name.  Coordinate-based names beginning with "F" are sources in the IRAS Faint Source Catalog.  Col.(2): Observation start date.  Col.(3): Whether the source is extended when compared with the telescope PSF.  The symbol ``\nodata'' denotes \textit{XMM-Newton} observations where radial profiles were not created; all these observations are assumed to be unresolved due to the large PSF of the telescope relative to \textit{Chandra}.   Col.(4): Total counts in the 0.5-8.0~keV band.  Col.(5): Total counts in the 0.5--2.0~keV band.  Col.(6): Total counts in the 2.0--8.0~keV band.  Col.(7): The hardness ratio.  Col.(8):  The estimated $\Gamma$ based on the hardness ratio and assuming Galactic absorption.  Col.(9):  The estimated total absorption assuming $\Gamma = 1.7$ in units of 10$^{22}$ cm$^{-2}$.  Col.(10):  The estimated observed 0.5--2.0~keV flux assuming the estimated $\Gamma$ and Galactic absorption in units of 10$^{-14}$ erg s$^{-1}$ cm$^{-2}$.  Col.(11):  The estimated observed 2.0--10.0~keV flux assuming the estimated $\Gamma$ and Galactic absorption in units of 10$^{-14}$ erg s$^{-1}$ cm$^{-2}$.  Col. (12): Derived 0.5--2~keV luminosity from Col. (10) in units of 10$^{42}$ erg s$^{-1}$.  Col. (13): Derived 2--10~keV luminosity from Col. (11) in units of 10$^{42}$ erg s$^{-1}$.
}
\label{tab:hr}
\end{deluxetable}

\begin{deluxetable}{clcccccccccccccccc}
\tablecolumns{18}
\rotate
\setlength{\tabcolsep}{0.02in}
\tabletypesize{\tiny}
\tablecaption{Best-fit Parameters for ULIRG Spectra with $>$ 100 Counts}
\tablewidth{0pt}
\tablehead{
\colhead{Galaxy} & \colhead{Date} &\colhead{Model}& \colhead{$kT$} & \colhead{$\Gamma$} & \colhead{$N_{\rm H, source 1}$} & \colhead{$f_{\rm cover 1}$}&\colhead{$N_{\rm H, source 2}$}&\colhead{$f_{\rm cover 2}$} & \colhead{$E_{\rm line}$}&\colhead{$\sigma$}&\colhead{EW}&\colhead{Stat}&\colhead{d.o.f}&\colhead{$F_{\rm 0.5-2 keV}$}&\colhead{$F_{\rm 2-10 keV}$}&\colhead{$L_{\rm 0.5-2 keV}^{corr}$}&\colhead{$L_{\rm 2-10 keV}^{corr}$}\\
\colhead{(1)} & \colhead{(2)} & \colhead{(3)} & \colhead{(4)} & \colhead{(5)} & \colhead{(6)} & \colhead{(7)} &\colhead{(8)} & \colhead{(9)} & \colhead{(10)} & \colhead{(11)} & \colhead{(12)} & \colhead{(13)} & \colhead{(14)}& \colhead{(15)} & \colhead{(16)} & \colhead{(17)} &\colhead{(18)}}
\startdata
\cutinhead{Cash Statistics (unbinned spectra with $<$ 1000 counts)}
          F00188$-$0856  
                 & 2004 Dec 20 & A  &  \nodata& 1.76$^{+0.33}_{-0.29}$  & \nodata&\nodata&\nodata&\nodata&\nodata&\nodata&\nodata&716.0&1940&0.42$^{+0.16}_{-0.18}$&0.77$^{+0.53}_{-0.63}$&0.20&0.33\\
                 F04103$-$2838\tablenotemark{a}  
                  & 2006 Feb 13& C&   0.10$^{+0.03}_{-0.08}$ &1.36$^{+0.97}_{-0.44}$& 0.19$^{+0.33}_{-0.19}$&\nodata&\nodata&\nodata&6.43$^{+0.26}_{-0.26}$&0.26$^{+0.24}_{-0.26}$&1.62$^{+1.58}_{-1.12}$&848.1&1953&2.17$^{+0.87}_{-0.87}$&4.56$^{+2.04}_{-0.97}$&0.64&1.35 \\
                 F05024$-$1941  & 2007 Feb 8 &D&  \nodata&  1.94$^{+0.56}_{-0.47}$&32.9$^{+243.9}_{-29.5}$&0.86$^{+0.09}_{-0.40}$&\nodata&\nodata&\nodata&\nodata&\nodata&646.5&1938& 0.50$^{+0.38}_{-0.50}$& 2.28$^{+2.10}_{-2.28}$&4.45&5.62  \\
\enddata

\tablenotetext{a}{Values taken from \citet{teng08}.  The energy bands in that paper are 0.2--2~keV and 2--10~keV.}
\tablenotetext{b}{The ratio of the normalization parameters between the scattered and direct components is 249 for F15250+3508, 62 for F1119+3257, 24  and 18 for the 2000 and 2002 observation of Mrk~273, respectively.}
\tablenotetext{c}{The addition of an emission line at 6.4~keV does improve the fit, but not statistically significant.}
\tablenotetext{d}{The addition of a red-shifted oxygen edge improves the fit significantly.}
\tablenotetext{e}{The addition of this absorption component does improve the fit, but not significantly.}
\tablecomments{
Col.(1): Galaxy name.  Coordinate-based names beginning with "F" are sources in the IRAS Faint Source Catalog.  Col.(2): Observation start date.  Col.(3): Best-fit continuum model, all modified by Galactic absorption: A = single power law, B = MEKAL only, C = MEKAL plus power law, D = scattering model or partial covering model, E = blackbody plus power law.  Col.(4): MEKAL temperature, if applicable, in units of keV.  Col.(5): Spectral index of the power law (or reflection) model.  Col. (6)--(9): Absorption within the source and corresponding covering fraction for the covering fraction absorber model.  The absorption values are in units of $10^{22}$~cm$^{-2}$.  Col.(10): Rest energy of the emission line(s) in units of keV.  Col. (11): Line width(s) of the emssion line(s) in units of keV.  Col. (12): Equivalent width(s) of the emission line(s) in units of keV.  Col. (13)--(14):  Fitting statiscs and degrees of freedom, respectively.  Col. (15)--(16): Observed flux after background subtraction in units of 10$^{-14}$~erg~s$^{-1}$~cm$^{-2}$.  Col. (17)--(18): Nominal absorption corrected luminosities in units of 10$^{42}$~erg~s$^{-1}$. 
}

\label{tab:ulirgfits}
\end{deluxetable}

\begin{deluxetable}{cccc}
\tablecolumns{13}
\setlength{\tabcolsep}{0.03in}
\tabletypesize{\scriptsize}
\tablecaption{PG QSOs: Tested Universal Spectral Models}
\tablewidth{0pt}
\tablehead{
  \colhead{ID}&\colhead{AGN}&\colhead{SE}&\colhead{$\chi^2_\nu$}\\
\colhead{(1)} & \colhead{(2)} & \colhead{(3)} & \colhead{(4)}}
\startdata
A&PL &\nodata&8.0\\
B&PL + Abs$\times$PL&\nodata&3.0\\
C&PCF$\times$PL&\nodata&3.4\\
D&PCF$\times$PCF$\times$PL&\nodata&1.8\\
E&PCF$\times$PCF$\times$PL&MEKAL&1.5\\
F&PL&CompTT&2.1\\
G&PL&REFLION (blurred; C06)& 19.9\\
H&PL&PEXRAV& 3.6\\
I&PL&PEXRIV& 3.1\\
J&PL+PEXRAV&zBbody&1.3\\
K&PL+PEXRIV&zBbody&1.2\\
L&C06&zBbody&1.4\\
M&PL+PEXRIV&zBbody+Zedge&1.1\\
\enddata
\tablecomments{
Col.(1): Model identifier.  Col.(2): Spectral models fitting the AGN component: PL = redshifted power law, Abs$\times$PL = absorbed redshifted power law, PCF = partial covering absorber.  Col.(3): Spectral models fitting the soft excess (SE) component: CompTT = Comptonization model, REFLION= blurred ionized reflection model of C06, PEXRAV = neutral reflection, PEXRIV = ionized reflection, ZBbody = redshifted blackbody, Zedge = redshifted absorption edge. (4): Reduced $\chi^2$ from the fits.    
}\label{tab:modstat}
\end{deluxetable}

\begin{deluxetable}{ccc|cc|cccc|cccc}
\tablecolumns{13}
\setlength{\tabcolsep}{0.03in}
\tabletypesize{\scriptsize}
\tablecaption{PG~QSOs: Derived Parameters from Model M}
\tablewidth{0pt}
\tablehead{
\colhead{}&\colhead{}&\colhead{}&\multicolumn{2}{c}{Soft Excess}&\multicolumn{4}{c}{Power Law}&\multicolumn{4}{c}{Reflection}\\
\colhead{Galaxy} & \colhead{$N_H$} &\colhead{$\xi$}&\colhead{$F_{0.5-2}$} &\colhead{$L_{0.5-2}$} &\colhead{$F_{0.5-2}$} &\colhead{$F_{2-10}$} &\colhead{$L_{0.5-2}$} & \colhead{$L_{2-10}$}&\colhead{$F_{0.5-2}$} &\colhead{$F_{2-10}$} &\colhead{$L_{0.5-2}$} & \colhead{$L_{2-10}$}\\
\colhead{} & \colhead{} & \colhead{} & \colhead{($\times 10^{-14}$)} & \colhead{($\times 10^{42}$)} & \colhead{($\times 10^{-12}$)} & \colhead{($\times 10^{-12}$)} &\colhead{($\times 10^{44}$)} &\colhead{($\times 10^{44}$)} &\colhead{($\times 10^{-14}$)}&\colhead{($\times 10^{-14}$)} & \colhead{($\times 10^{44}$)} & \colhead{($\times 10^{44}$)}\\
\colhead{(1)} & \colhead{(2)} & \colhead{(3)} & \colhead{(4)} & \colhead{(5)} & \colhead{(6)} & \colhead{(7)} &\colhead{(8)} &\colhead{(9)} &\colhead{(10)}&\colhead{(11)} & \colhead{(12)} & \colhead{(13)}}
\startdata
                  PG0050$+$124  &0.02 &0.0&  0.81& 0.07&7.92&5.74&0.71&0.51&0.05&1.05&0.005&0.09\\
                  PG0838$+$770  & 0.78 &103.9& 16.32&7.39&0.74&0.50&0.34&0.22&0.01&0.11&0.006&0.05\\
                  PG0844$+$349  & 0.00&0.0& 128.93  & 12.75&6.15&4.12&0.61&0.41&0.07&1.29&0.007&0.13\\
                  PG0953$+$414  & 0.00&0.0  & 25.62  &41.88 &3.25&2.17&5.32&3.55&0.05&0.86&0.08&1.41\\
                  PG1001$+$054  & 3.03&155.2& 1.39&0.99&0.08&0.05&0.06&0.04&0.01&0.07&0.006&0.05\\
                  PG1004$+$130  & 0.56&12.0 &  1.62&2.81  &0.19&0.13&0.33&0.22&0.01&0.17&0.02&0.30\\
                  PG1116$+$215  & 0.00&0.1 & 43.18  &37.38&3.80&2.54&3.29&2.20&0.05&0.87&0.04&0.75\\
                  PG1126$-$041  & 4.16&1.3  & 3.71 & 0.32&1.40&0.94&0.12&0.08&0.03&0.49&0.002&0.04\\
\enddata
\tablecomments{
Col.(1): Galaxy name.  Col.(2): Intrinsic column density in units of 10$^{22}$ cm$^{-2}$.  Col.(3): Ionization parameter from the PEXRIV component in erg cm s$^{-1}$.  Col.(4): Absorption-corrected 0.5--2~keV flux for the blackbody component in units of 10$^{-14}$~erg~s$^{-1}$~cm$^{-2}$.  The 2--10~keV flux of this component is $\sim$10$^3$ smaller, and so is insignificant .  Col.(5): 0.5--2~keV luminosity in units of 10$^{42}$~erg~s$^{-1}$ for the blackbody component.  Col.(6)--(7):  Absorption-corrected 0.5--2~keV and 2--10~keV flux for the power law component in units of 10$^{-12}$~erg~s$^{-1}$~cm$^{-2}$.  Col.(8)--(9): Absorption corrected 0.5--2~keV and 2--10~keV luminosity of the power law component in units of 10$^{44}$~erg~s$^{-1}$.  Col.(10)--(11):  Absorption-corrected 0.5--2~keV and 2--10~keV flux in units of $10^{-12}$~erg~s$^{-1}$~cm$^{-2}$ for the reflected component.  Col.(12)--(13): Absorption-corrected 0.5--2~keV and 2--10~keV luminosity of the reflected component in units of $10^{44}$~erg~s$^{-1}$.
}
\label{tab:pexrivstat}
\end{deluxetable}

\begin{deluxetable}{cccccccccc}
\tablecolumns{10}
\setlength{\tabcolsep}{0.03in}
\tabletypesize{\scriptsize}
\tablecaption{Quantifying the Errors in the Hardness Ratio Method}
\tablewidth{0pt}
\tablehead{
\colhead{Exposure} &\colhead{Hard} & \colhead{Soft} & \colhead{HR} &\colhead{$\Gamma$} &\colhead{N$_{H}$}&\colhead{$F_{0.5-2~keV}$}&\colhead{\% error$_{0.5-2~keV}$}&\colhead{$F_{2-10~keV}$}&\colhead{\% error$_{2-10~keV}$}\\
\colhead{(1)} & \colhead{(2)} & \colhead{(3)} & \colhead{(4)} & \colhead{(5)} & \colhead{(6)} & \colhead{(7)}& \colhead{(8)}&\colhead{(9)}&\colhead{(10)}}
\startdata
Input&\nodata&\nodata&\nodata&1.8&\nodata&1.77&\nodata&2.95&\nodata\\
\hline
100&139.6$\pm$11.7&453.2$\pm$21.2&--0.53$\pm$0.05&1.80$^{+0.13}_{-0.12}$&\nodata&1.77&0.0&2.94&--0.3\\
75&104.7$\pm$10.6&339.3$\pm$18.3&--0.53$\pm$0.05&1.80$^{+0.14}_{-0.13}$&\nodata&1.77&0.0&2.94&--0.3\\
50&70.1$\pm$8.1&225.6$\pm$15.3&--0.53$\pm$0.07&1.80$^{+0.17}_{-0.15}$&\nodata&1.76&--0.6&2.93&--0.7\\
25&35.0$\pm$5.9&113.3$\pm$10.6&--0.53$\pm$0.09&1.80$^{+0.25}_{-0.21}$&\nodata&1.77&0.0&2.94&--0.3\\
20&27.9$\pm$5.1&90.1$\pm$9.5&--0.53$\pm$0.10&1.80$^{+0.28}_{-0.24}$&\nodata&1.76&--0.6&2.92&--1.0\\
15&21.0$\pm$4.6&67.8$\pm$8.3&--0.53$\pm$0.12&1.80$^{+0.33}_{-0.27}$&\nodata&1.77&0.0&2.93&--0.7\\
10&14.0$\pm$3.8&45.2$\pm$6.6&--0.53$\pm$0.15&1.80$^{+0.41}_{-0.32}$&\nodata&1.77&0.0&2.93&--0.7\\
\enddata

\tablecomments{
Col.(1): Simulated exposure time in ksec.  Col.(2): Average counts in the hard band from 1000 simulations.  Col.(3): Average counts in the soft band from 1000 simulations.  Col.(4): The hardness ratio.  Col.(5): Spectral index derived from the hardness ratio method.  Col.(6): Input column density from within the source in units of cm$^{-2}$.  Col.(7): Observed 0.5--2~keV flux in units of $10^{-14}$ erg s$^{-1}$ cm$^{-2}$ assuming only Galactic absorption.  Col.(8): The percent error of the simulated 0.5--2~keV flux relative to that of the input model.   Col.(9):  Same as (7), but for the 2--10~keV band.  Col.(10):  Same as (8), but for the 2--10~keV band.
}
\label{tab:hrsims}
\end{deluxetable}


\begin{figure}
\figurenum{1}
\epsscale{1}
\plotone{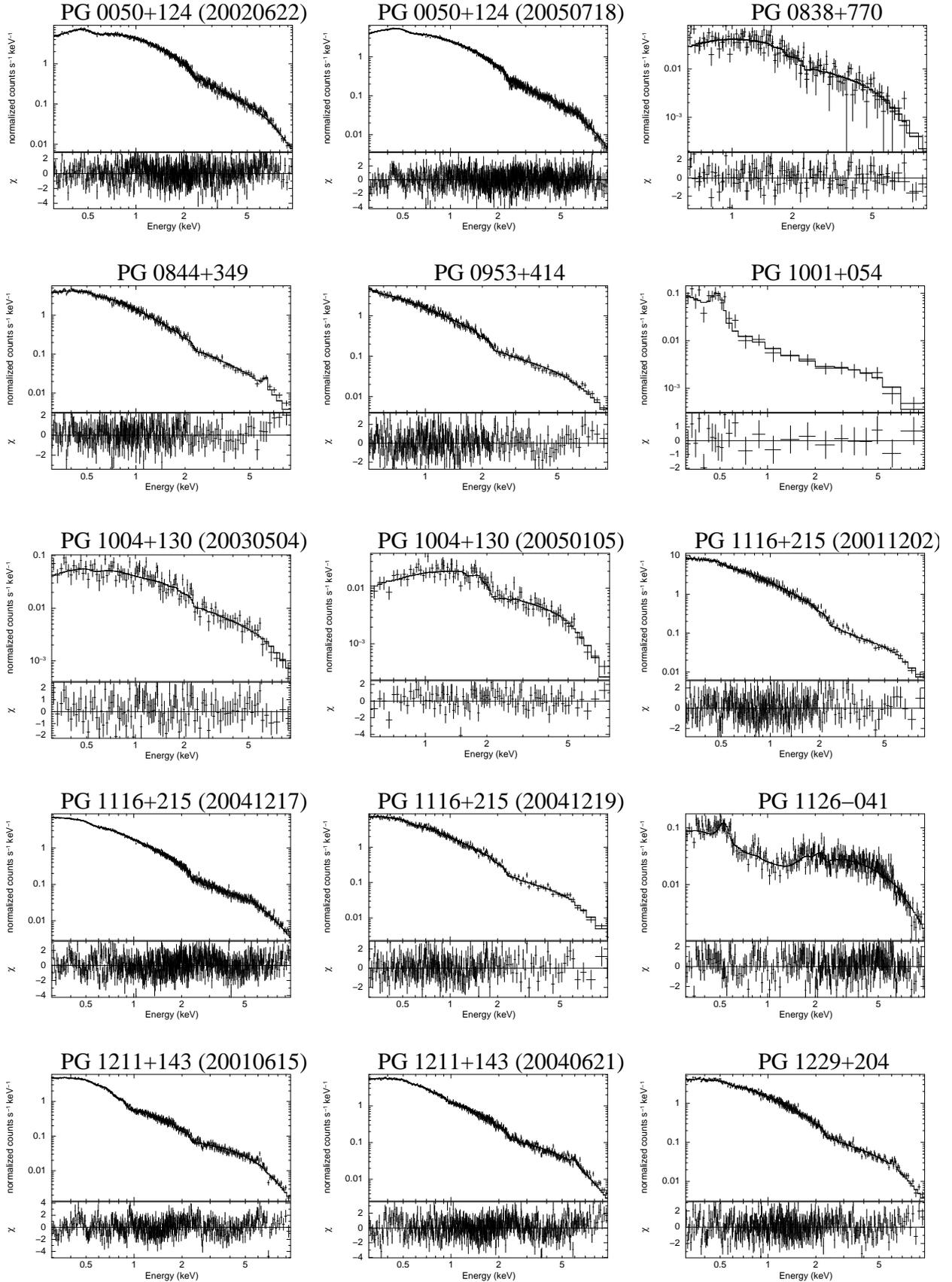}
\caption{The spectra of the PG QSOs with the best-fit power-law model.}
\label{fig:allspecs}
\end{figure}

\begin{figure}
\figurenum{1}
\epsscale{1}
\plotone{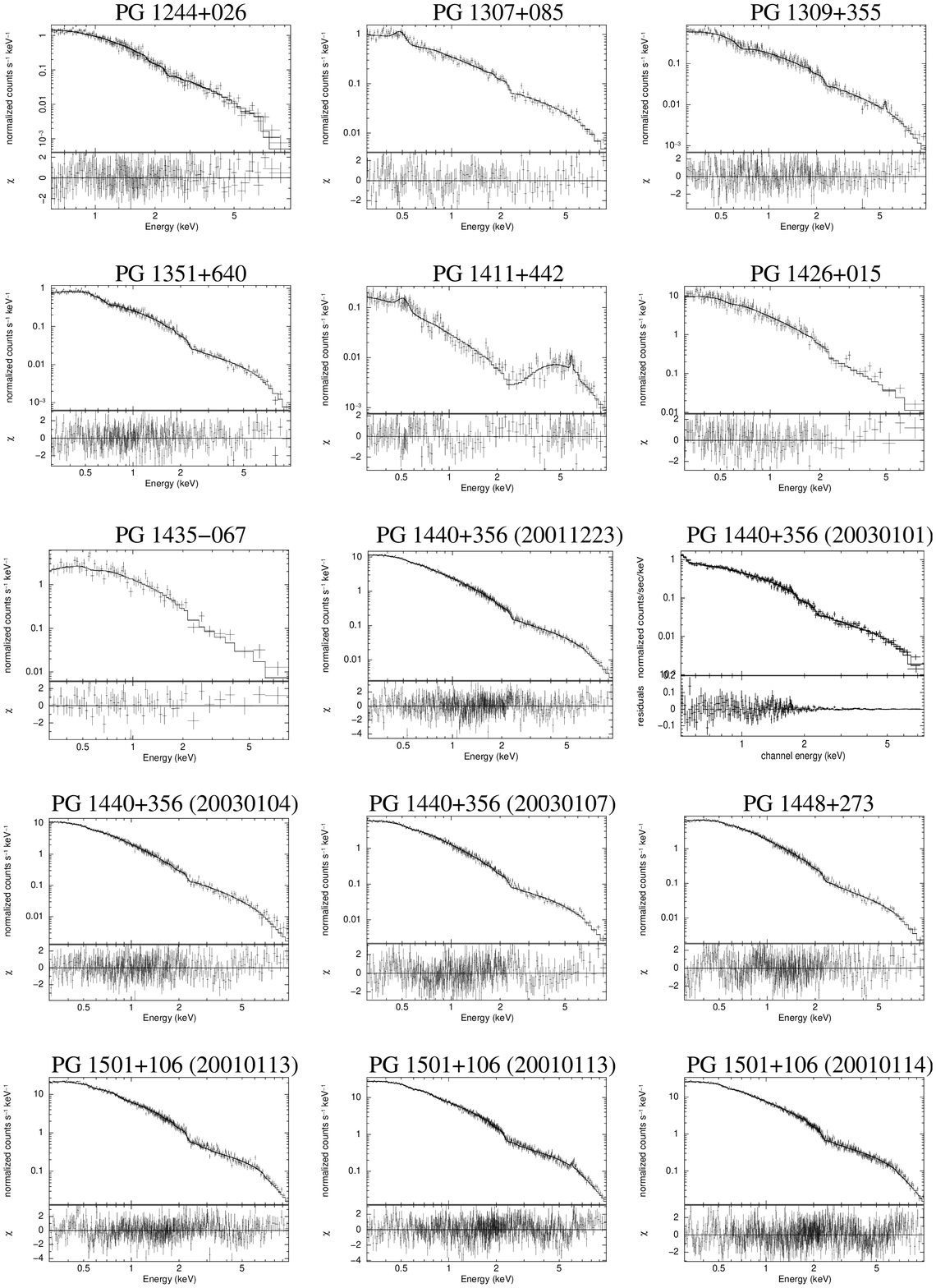}
\caption{cont.}
\end{figure}

\begin{figure}
\figurenum{1}
\epsscale{1}
\plotone{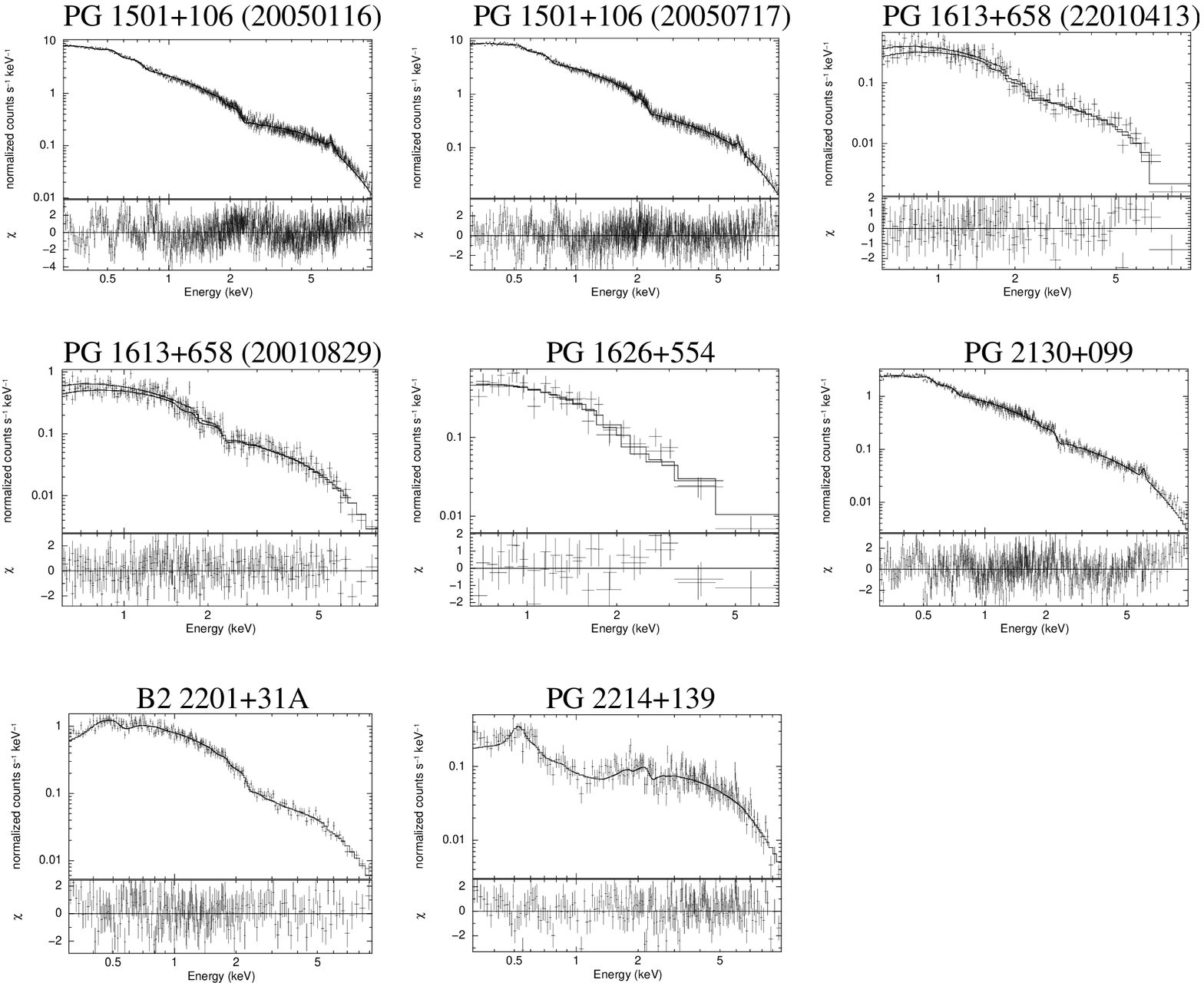}
\caption{cont.}
\end{figure}

\begin{figure}
\figurenum{2}
\epsscale{.8}
\plotone{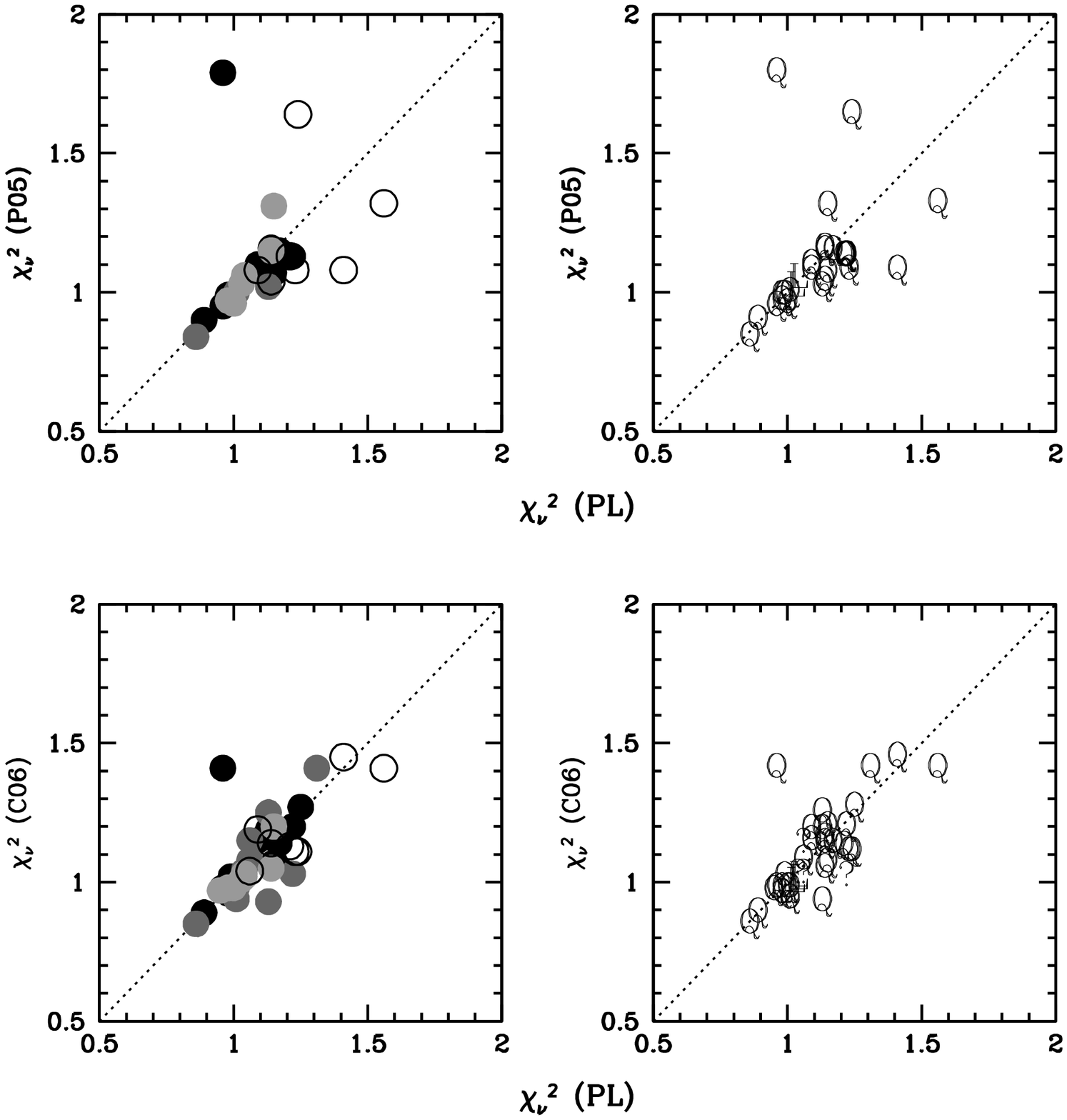}
\caption{Comparison of the reduced $\chi^2$ values from the power-law
models with those of P05 (top) and C06 (bottom) for
all PG~QSOs.  In the left panels, the symbols represent the
\citet{netzer} FIR SED classification of each source where
black circles represent strong FIR emitters, dark gray circles weak
FIR emitters, light gray circles undetected FIR emitters, and open
circles sources with unknown SED classifications.  In the right
panels, the symbols represent radio loudness.  With the exception of
two sources (PG~1211+143 and PG~1501+106), the statistics suggest that
the power-law models are as good, or better, descriptions of the
quasar spectra as the P06 and C06 models.}
\label{fig:chicompp}
\label{fig:chicompc}
\end{figure}

\begin{figure}
\figurenum{3}
\epsscale{1}
\plotone{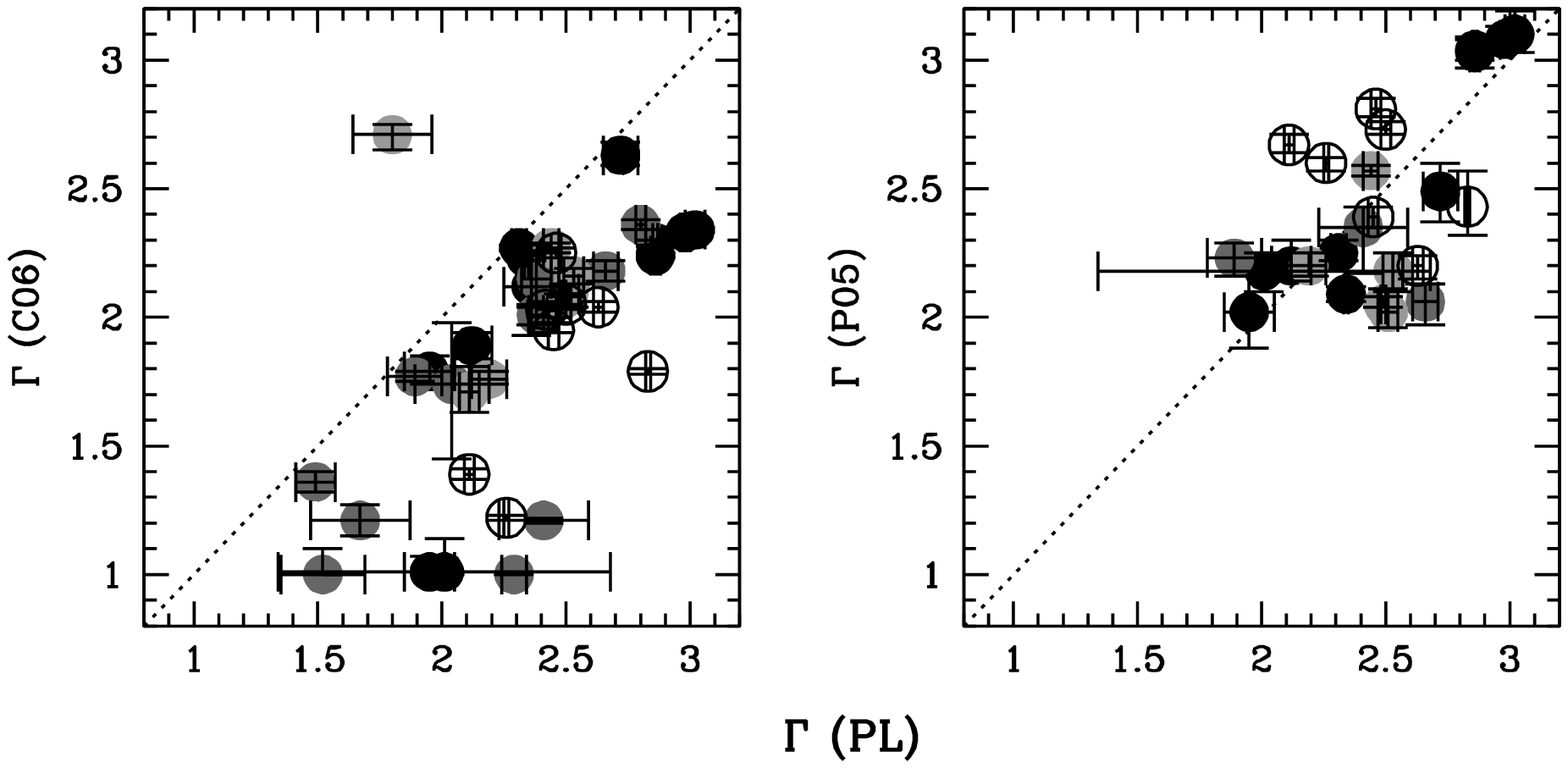}
\caption{Comparison of the spectral index for the power-law models with those from the C06 (left panels) and P05 (right panels) models.  The meaning of the symbols is the same as that in Figure~\ref{fig:chicompp} for the SED classifications.  The dotted lines are one-to-one ratios to help guide the eye.  The indices based on the power-law and the P05 models tend to be steeper than those from the C06 model.  Almost all of the indices are softer than the canonical value of 1.8 for AGNs but are still within the upper end of the range observed in other AGNs.  Because the P05 model is not universal for the PG~QSOs, the P05 figure is missing data from sources in their sample that do not overlap with our sample.}
\label{fig:gammacomp}
\end{figure}

\clearpage

\begin{figure}
\figurenum{4}
\epsscale{.7}
\plotone{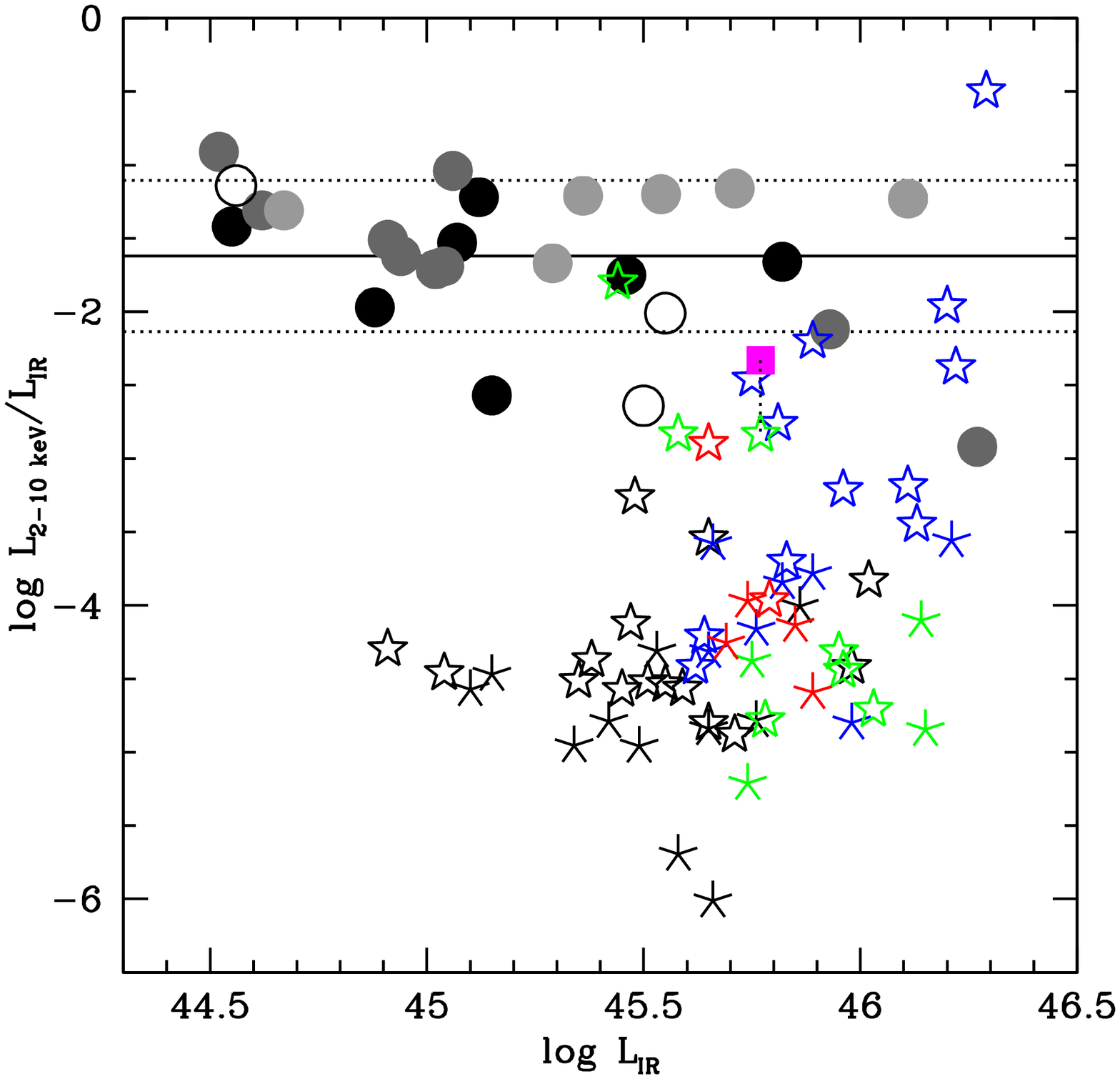}
\caption{Comparison of 2--10~keV luminosity and the IR luminosity for
the PG~QSOs and the U/LIRGs.  The horizontal axis is the IR luminosity and the
vertical axis is the absorption-corrected 2--10~keV luminosity divided by the
IR luminosity.  The symbols for the PG~QSOs are the same as those in
Figure~\ref{fig:chicompp} for the SED and the values are from the
global modeling in \S~\ref{sec:prefmod}.  For the U/LIRGs, the open
stars represent the values derived from spectral fitting and the
skeletal stars represent the values derived from the HR method.  The
colors of the stars symbolize the effective optical depth
($\tau_{eff}$) taken from V09a where red represents objects
with the highest $\tau_{eff}$, green intermediate $\tau_{eff}$, and
blue the lowest $\tau_{eff}$.  The black stars are values from the
RBGS archive sample (median $L_{IR} = 10^{11.98} L_\odot$), where
$\tau_{eff}$ is unavailable.  The magenta square is the
absorption-corrected value for Mrk~273 derived from \textit{Suzaku}
data \citep{teng09} linked with the value derived from only the
\textit{Chandra}/\textit{XMM-Newton} data to demonstrate the effects
of improved absorption correction.  The solid line is the average
$\log$ ($L_{2-10 keV}/L_{IR}$) for the PG~QSOs ($\sim$ --1.6) and the
dotted lines represent 1-$\sigma$.  The $\log$ hard X-ray-to-IR
luminosity ratios of the U/LIRGs span about three orders of magnitude.}
\label{fig:lircomp}
\end{figure}

\begin{figure}
\figurenum{5}
\epsscale{.8}
\plotone{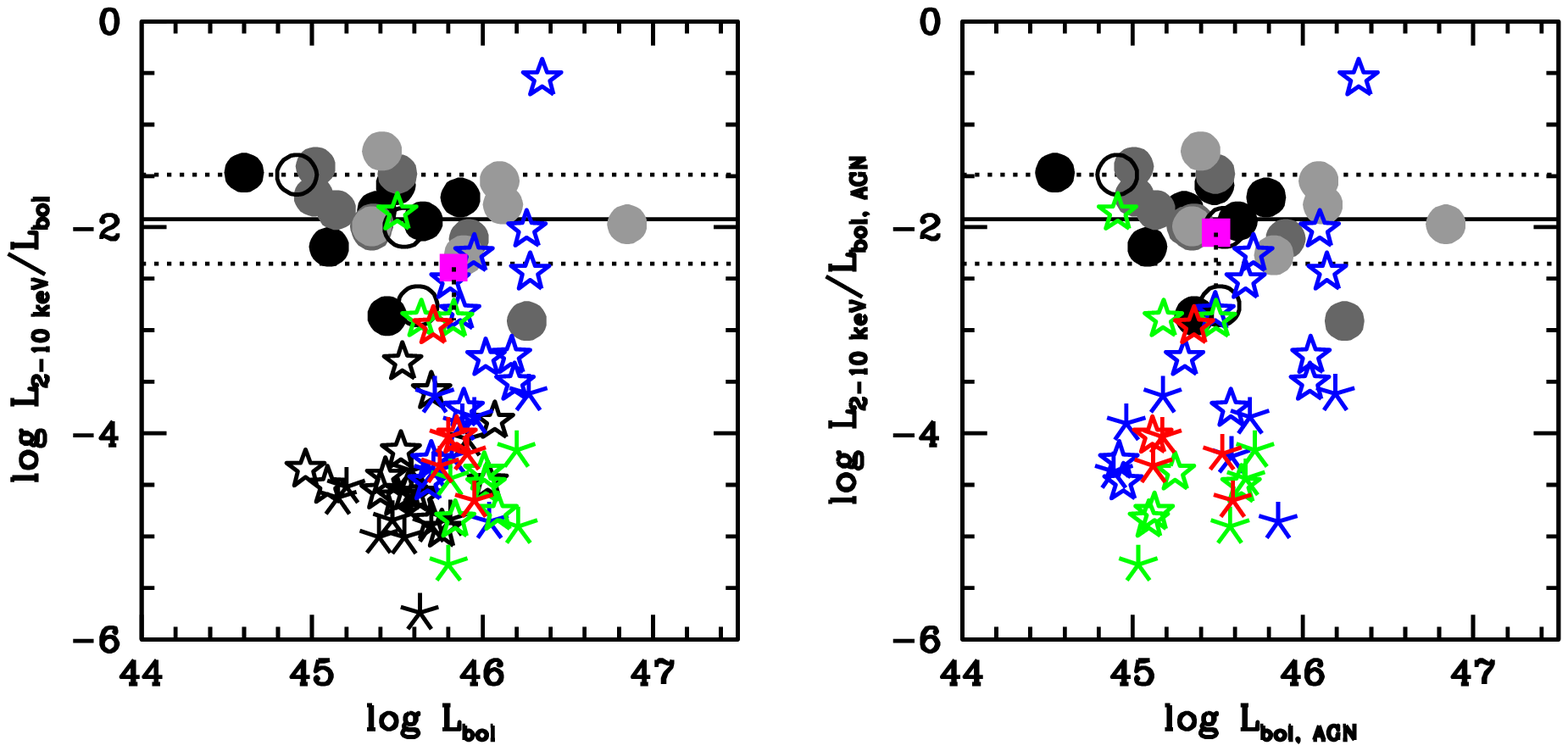}
\caption{Comparison of the bolometric luminosity from V09a
  with the absorption-corrected 2--10~keV luminosity.  The symbols for
  the PG~QSOs are the same as those in Figure~\ref{fig:chicompp} and
  the symbols for the U/LIRGs are the same as those in
  Figure~\ref{fig:lircomp}.  In the left figure, the horizontal axis is the
  total bolometric luminosity.  For the PG~QSOs, the 2--10~keV to bolometric luminosity ratio is nearly constant across the full range of bolometric luminosities. 
  The solid line is the average 2--10~keV to bolometric luminosity
  ratio for the PG~QSOs and the dotted lines represent one standard deviation of the mean.  The total 0.5--10~keV luminosity of the PG~QSOs are
  $\sim 0.5-11$\% of the bolometric luminosity.  Nearly all of the
  U/LIRG values fall below this trend.  The right panel plots the same
  2--10~keV luminosity as a function of the AGN bolometric luminosity
  ($L_{\rm bol, AGN} = f_{\rm AGN} L_{\rm bol}$) from V09a.
  The lines are the same as those in the left panel.  Once again, a
  tight range is seen among the PG~QSOs and most U/LIRGs fall
  below the relation.}
\label{fig:lbolcomp}
\end{figure}

\begin{figure}
\figurenum{6}
\epsscale{.7}
\plotone{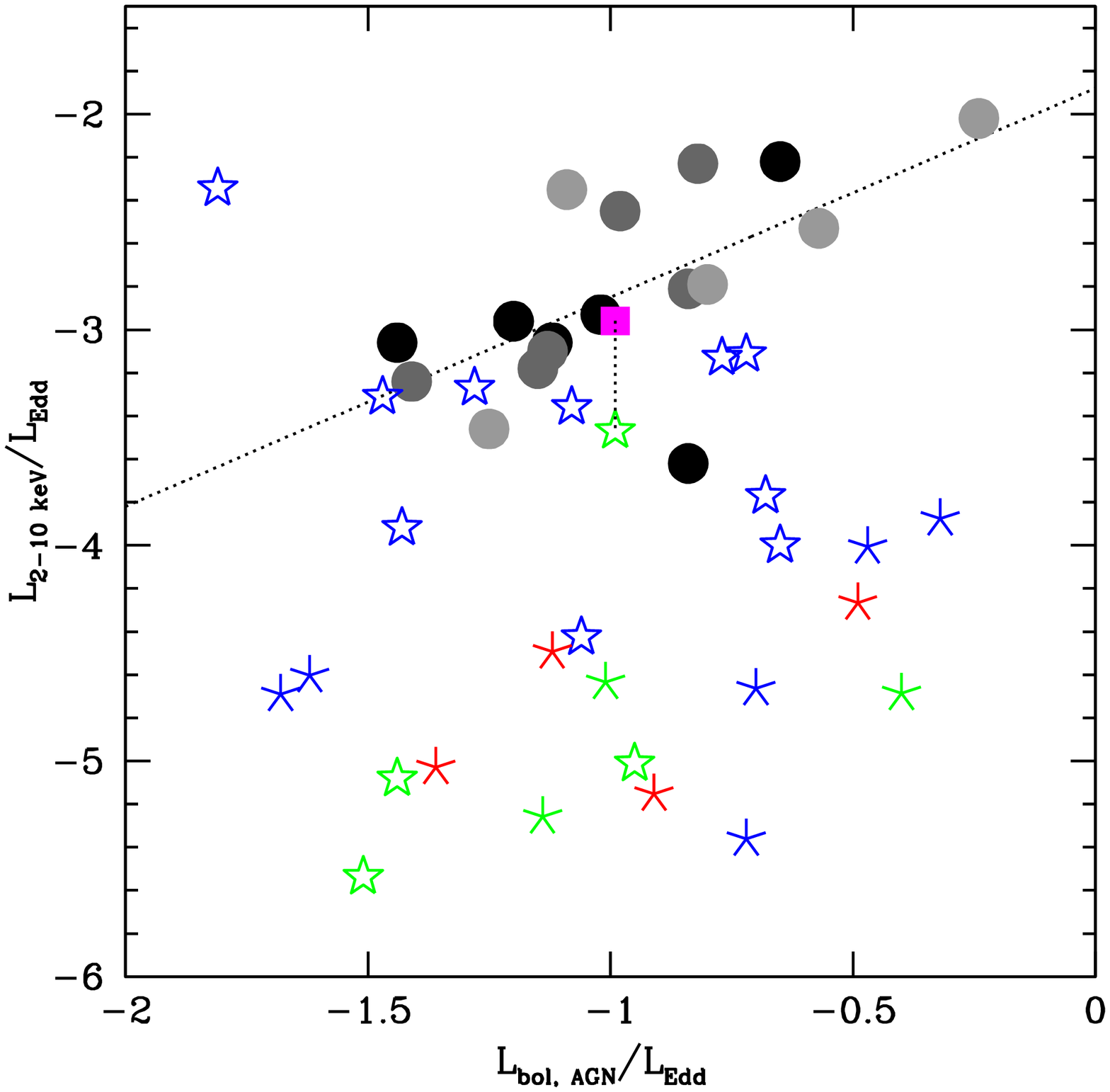}
\caption{We compare the X-ray and {\em Spitzer}-derived Eddington ratios for the U/LIRGs (same symbols as Figure~\ref{fig:lircomp}) and the PG~QSOs (same symbols as Figure~\ref{fig:chicompp}).  The black hole masses used to calculate the Eddington luminosity are taken from \citet{vei09b}.  The PG~QSO values determined by the two methods are linearly related.  
The dotted line represents this linear relationship where $\log ER_{X} = (0.97 \pm 0.52) \log ER_{IR} - (1.88 \pm 0.51)$ with $R^2 = 0.43$.  This correlation is significant at the 99.97\% confidence level.  Unlike the quasars, most U/LIRG values do not follow this relation. }
\label{fig:eddcomp}
\end{figure}

\begin{figure}
\figurenum{7}
\epsscale{.8}
\plotone{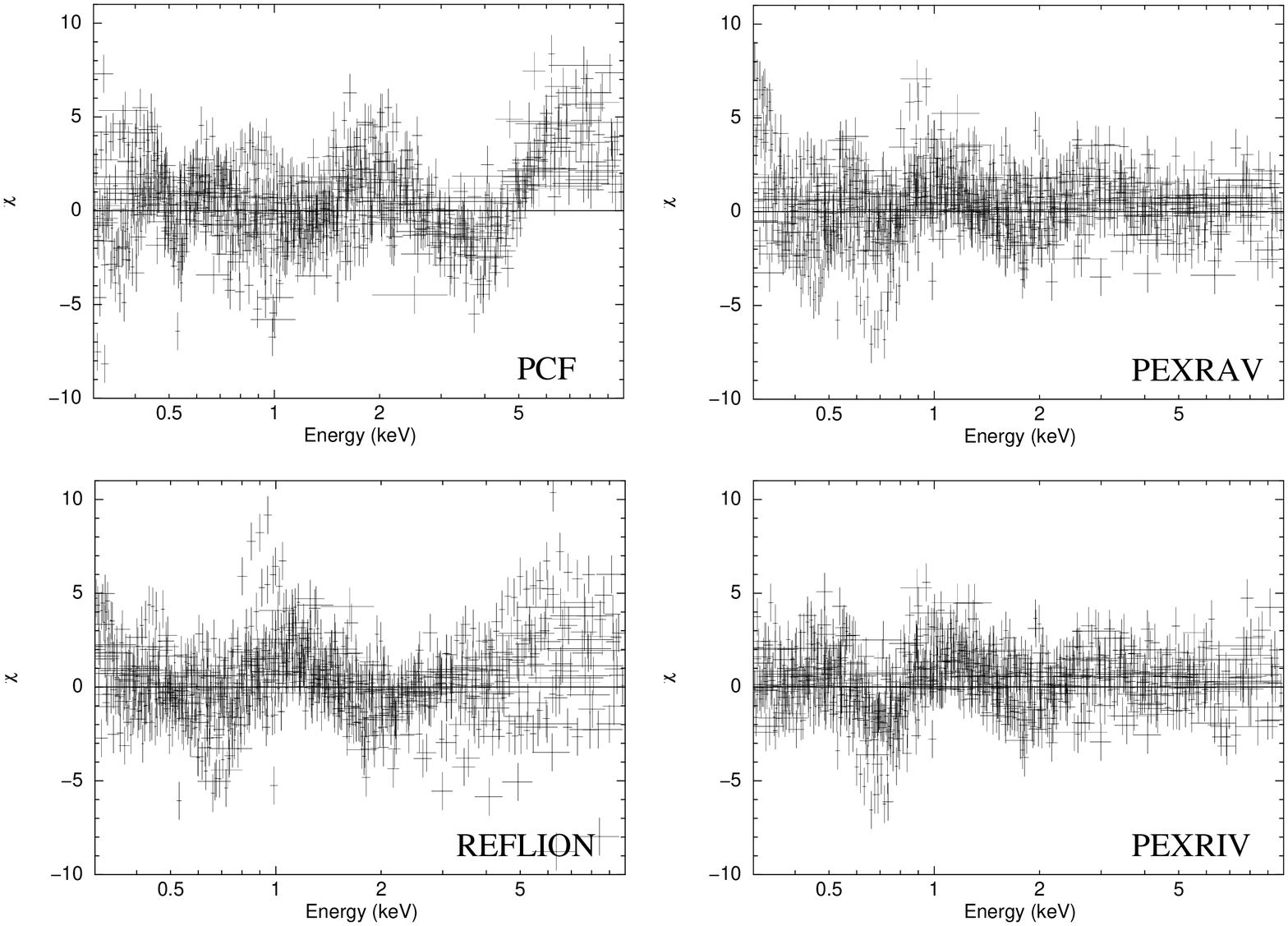}
\caption{Comparison of the goodness of fit between the different models for the PG~QSOs as a class.  The horizontal axis is energy in the observer's frame and the vertical axis is the residuals in terms of sigmas.  These are residuals for each spectrum plotted in the same frame to show, generally, where each model fails to describe the quasar spectra.  Four different models are compared for the \textit{continuum} of the X-ray spectrum.  The residuals are binned to at least 10-sigmas for display only.  Clockwise from the top left, the models presented are (1) MEKAL plus two partial covering absorbers (PCF; $\chi ^2 _\nu \sim 1.47$), (2) a redshifted blackbody plus a neutral reflection model (PEXRAV; $\chi ^2 _\nu \sim 1.29$), (3) a redshifted blackbody plus an ionized reflection model (PEXRIV; $\chi^2_\nu \sim 1.18$), and (4) a redshifted blackbody plus a more complex and blurred ionized reflection model (REFLION; $\chi ^2 _\nu \sim 1.38$).  All of the above models include absorption by the Galaxy and an underlying power-law model for the AGN component.  The PEXRIV model is the preferred model based on the fitting statistics and the residuals.  Note that there is an absorption feature between 0.5 and 0.7~keV in the observer's frame.  For the redshift range of our objects, this is consistent with the atomic transition of OVII and OVIII ($E \sim 0.7$~keV).}
\label{fig:allpndel}
\end{figure}

\begin{figure}
\figurenum{8}
\epsscale{.7}
\rotatebox{-90}{\plotone{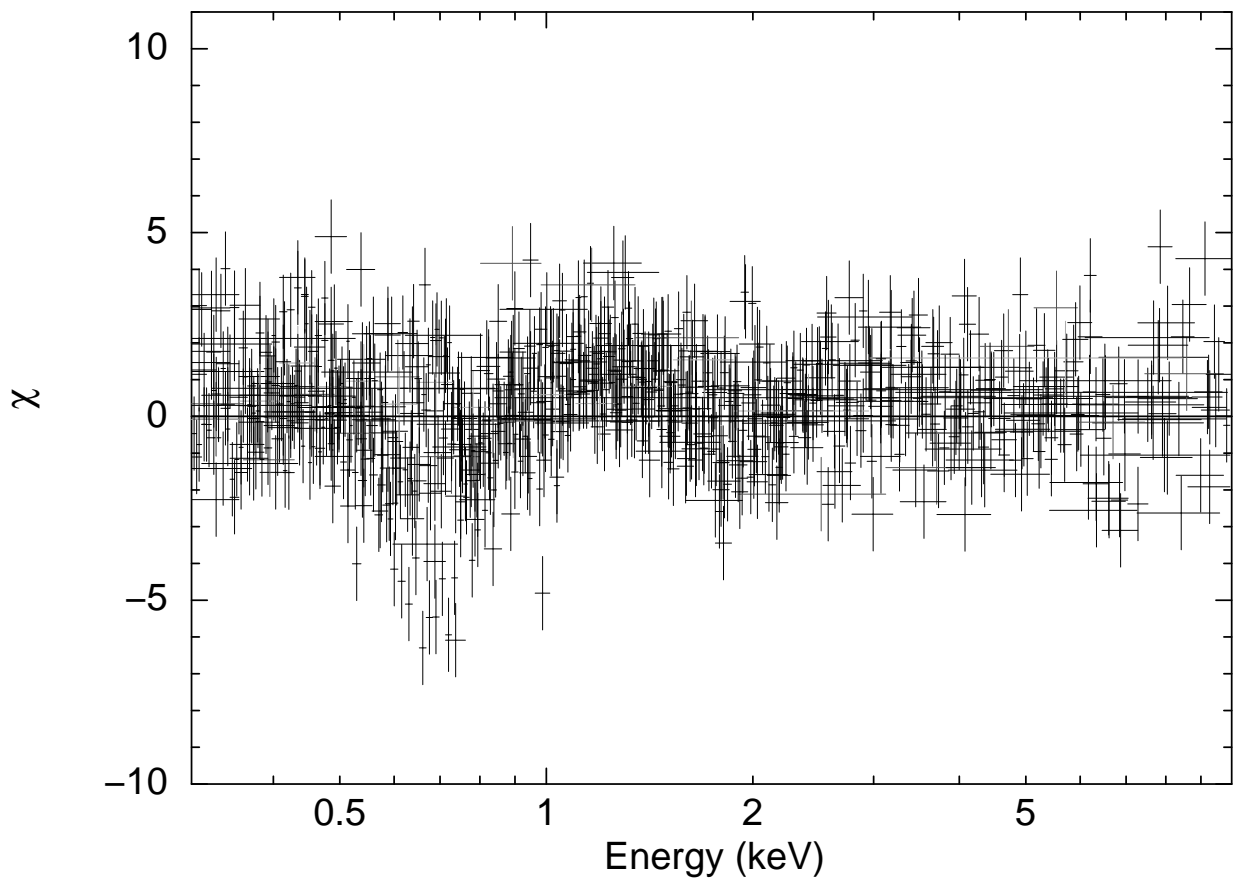}}
\caption{This is the same figure as the PEXRIV panel in the previous figure except with an additional redshifted absorption edge at $0.68^{+0.006}_{-0.003}$~keV with $\tau \sim 0.31 ^{+0.02}_{-0.03}$ ($\chi^2_\nu \sim 1.13$).  The absorption feature is consistent with the atomic transitions of O~VII or O~VIII.  The addition of the absorption edge component improves the fit significantly, with $\Delta \chi^2 \sim 338$ for a change in two degrees of freedom.  The large absorption feature still seen near 0.7~keV in the residuals is due to one source, PG~0050+124.}
\label{fig:pexrivdel}
\end{figure}

\begin{figure}
\figurenum{9}
\epsscale{.8}
\plotone{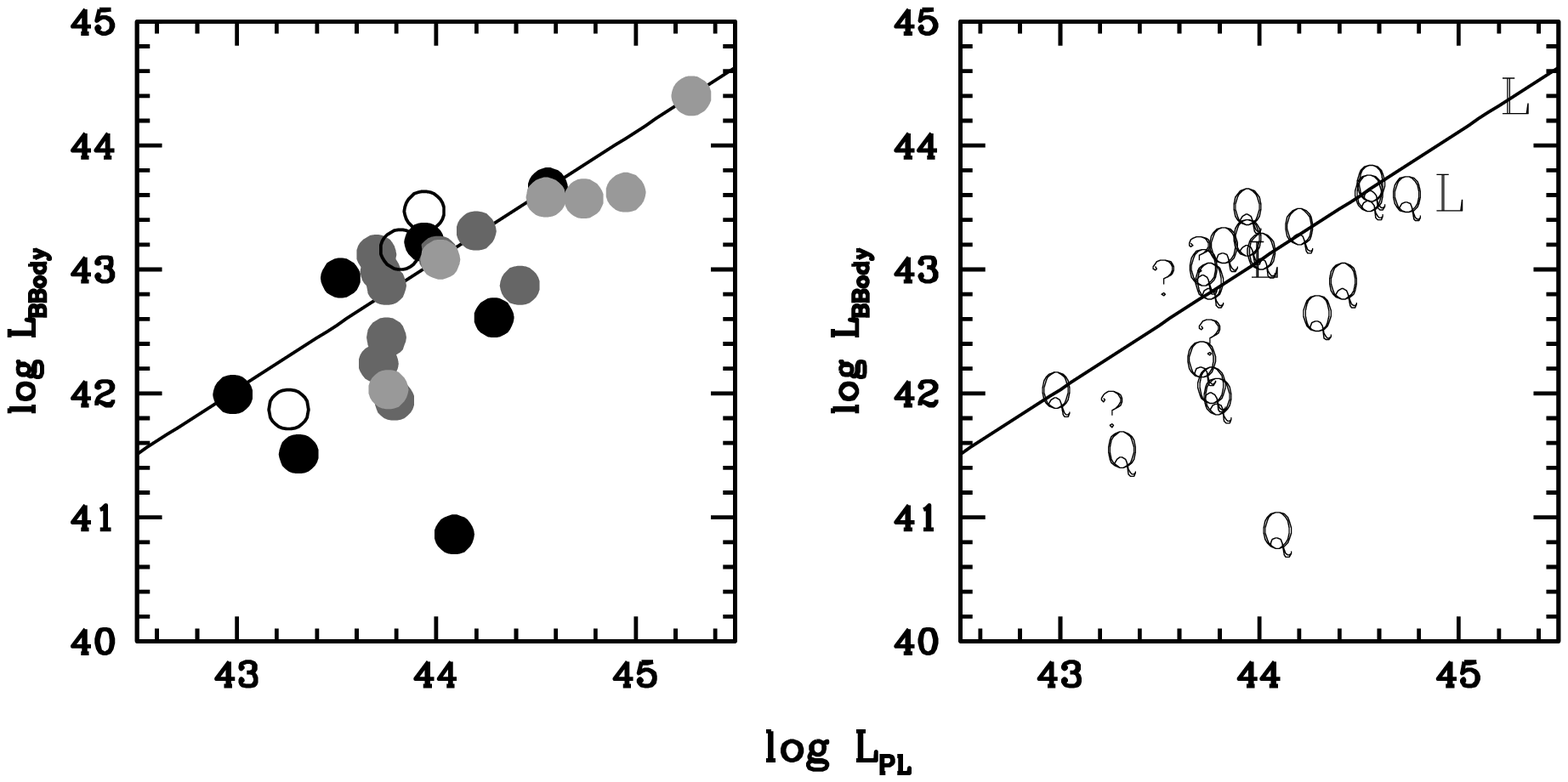}
\caption{Plots of the absorption-corrected blackbody (or,
  equivalently, the soft excess) luminosity versus power-law
  luminosity from the best-fit PEXRIV model to the PG~QSOs.  The meaning of the symbols is the same as that in
  Figure~\ref{fig:chicompp}.  The solid line is the
  linear regression fit to the data except for PG~0050+124, the
  outlier near (44, 41).  This source has an unusually strong absorption edge that the global model does not properly account for (see Figure~\ref{fig:pexrivdel}).  The correlation is significant ($>$99.99\%)
  with $R^2 = 0.65$ for $\log L_{\rm bbody} = (1.04 \pm 0.28) \log
  L_{\rm PL} - (2.69 \pm 12.28)$.  The slope of the line implies that
  $L_{\rm PL} \propto L_{\rm bbody}$.  This result suggests a direct physical link between the
  soft excess emission and the power-law component.}
\label{fig:ktvplref}
\end{figure}

\clearpage

\begin{figure}
\figurenum{10}
\epsscale{0.8}
\plotone{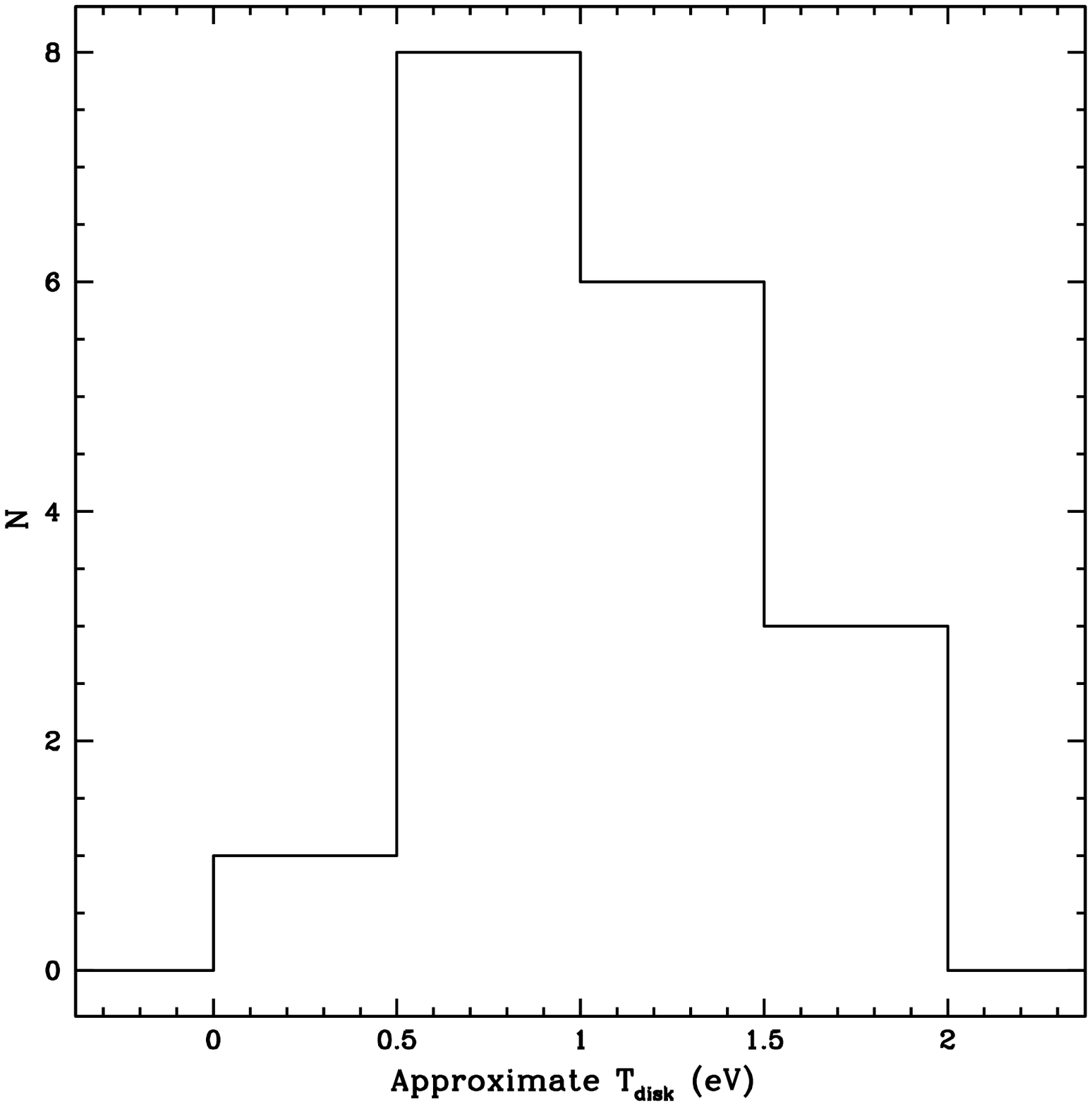}
\caption{A histogram of the thermal temperature of the accretion disk as derived from the photometric black hole mass of \citet {vei09b} and the 2--10~keV luminosity.  The scatter is small and the histogram peaks near 1~eV with a standard deviation of 0.4~eV.  These are only estimates of the temperature because the 2--10~keV to bolometric correction for AGNs is uncertain.  The predicted disk temperatures for the quasars are too low to explain the observed soft excess temperatures.}
\label{fig:temphist}
\end{figure}

\begin{figure}
\figurenum{11}
\epsscale{.7}
\plotone{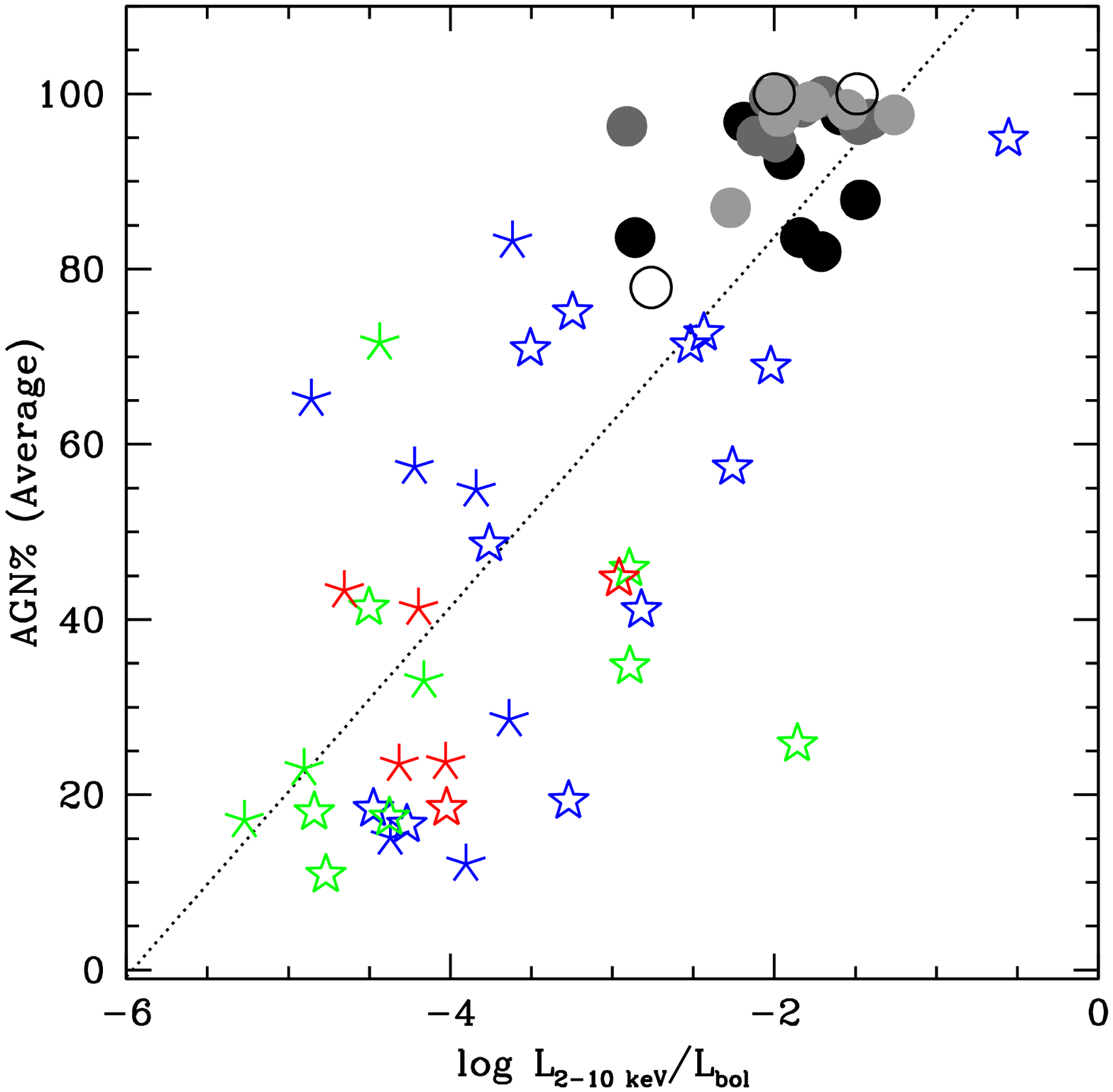}
\caption{AGN contribution to the bolometric luminosity of the QUEST ULIRGs (same symbols as Figure~\ref{fig:lircomp}) and PG~QSOs (same symbols as Figure~\ref{fig:chicompp}) derived by V09a against the logarithm of the 2--10~keV to bolometric luminosity ratio.  The AGN\% is the average of the six different methods (e.g., fine structure line ratios, PAH equivalent widths, mid-infrared continuum colors, and mid-infrared to far-infrared flux ratios) of measuring the AGN contribution by V09a.  The linear relationship (dotted line) is significant ($>$ 99.99\%) and  $R^2 = 0.64$: AGN\% $=(21.11 \pm 3.45) \log (L_{2-10 keV}/L_{bol}) + (125.89 \pm 11.04)$.  The correlation between the \textit{Spitzer}-derived AGN\% and the hard X-ray to bolometric luminosity ratio implies greater 2--10~keV flux with higher AGN contribution as one would expect.  The \textit{Spitzer}-derived AGN\% are likely more uncertain in objects with intermediate and high $\tau_{eff}$, but this does not seem to be the origin of the scatter in the relation. Instead we favor intrinsic variation in the 2-10 keV to bolometric luminosity ratio for pure AGNs or unsuspected obscuration of the hard X-rays. See discussion in \S~\ref{sec:methods} for more detail.}
\label{fig:agnpervedd}
\end{figure}

\clearpage

\begin{figure}
\figurenum{12}
\epsscale{.6}
\centering
\plotone{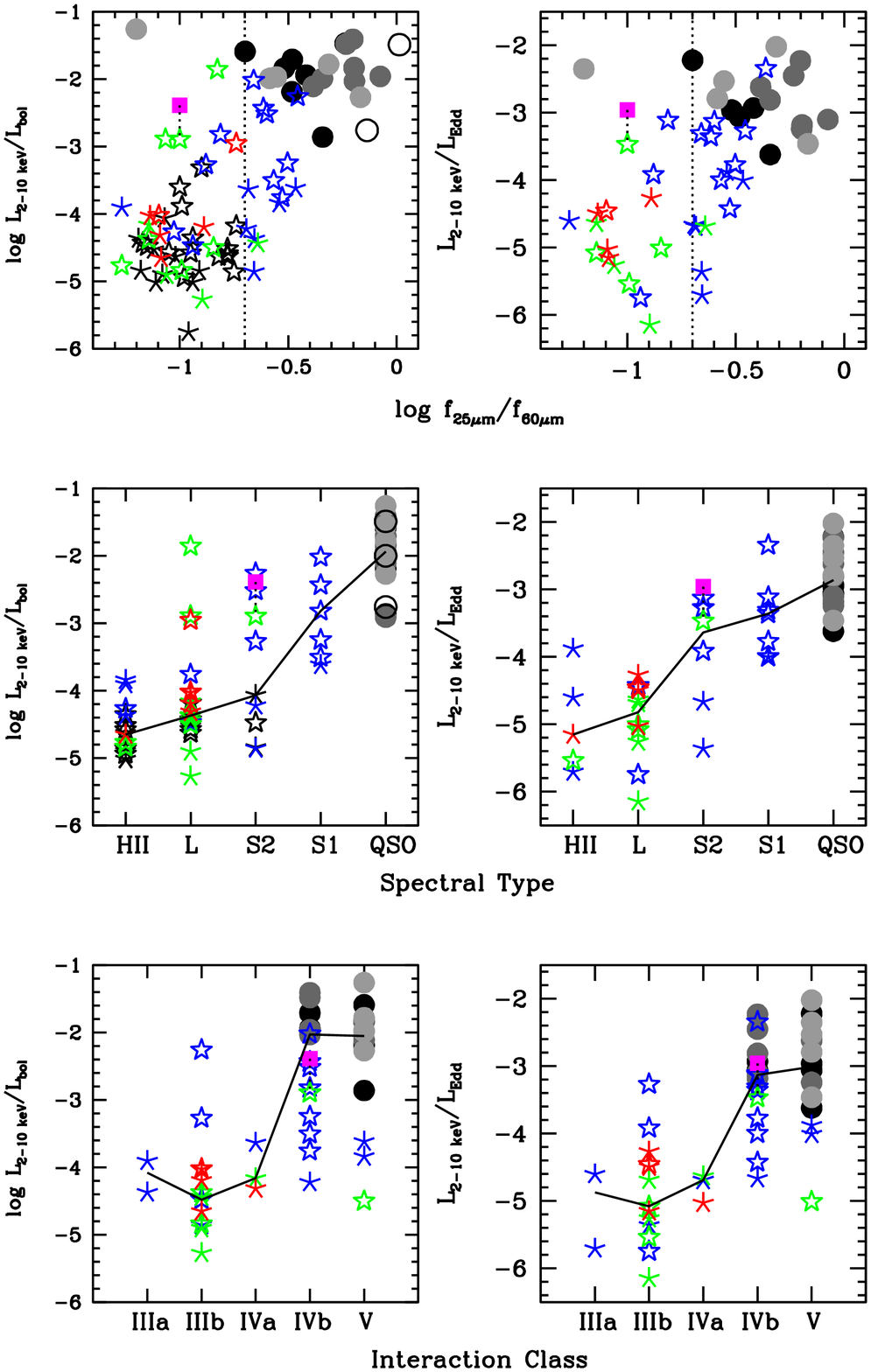}
\caption{A comparison of the 2--10~keV to bolometric luminosity ratio
  (our proxy for fractional AGN contribution to the bolometric
  luminosity; left) and the X-ray determined $\log$ ($L_{2-10
    keV}/L_{Edd}$) (right) with some of the key physical properties of
  the U/LIRGs and PG~QSOs: 25-to-60 $\mu$ dust temperature (top),
  optical spectral type (middle), and interaction class (bottom).  The
  symbols for the PG~QSOs are the same as those in
  Figure~\ref{fig:chicompp} for the SED and the symbols for the
  U/LIRGs are the same as those in Figure~\ref{fig:lircomp}.  In the
  bottom two rows, the line connects the median values for each
  type/class of objects.  The infrared-warmer objects have distinctly
  higher hard X-ray to bolometric luminosity ratios.  The more
  Seyfert-like ULIRGs and the more advanced mergers (IVb and V) also
  tend to have a stronger AGN component, and the PG~QSOs extend these
  trends.  These trends are similar to those found in the analysis of
  the \textit{Spitzer} data by V09a.}
\label{fig:propcomp}
\end{figure}

\begin{figure}
\figurenum{13}
\epsscale{.7}
\centering
\plotone{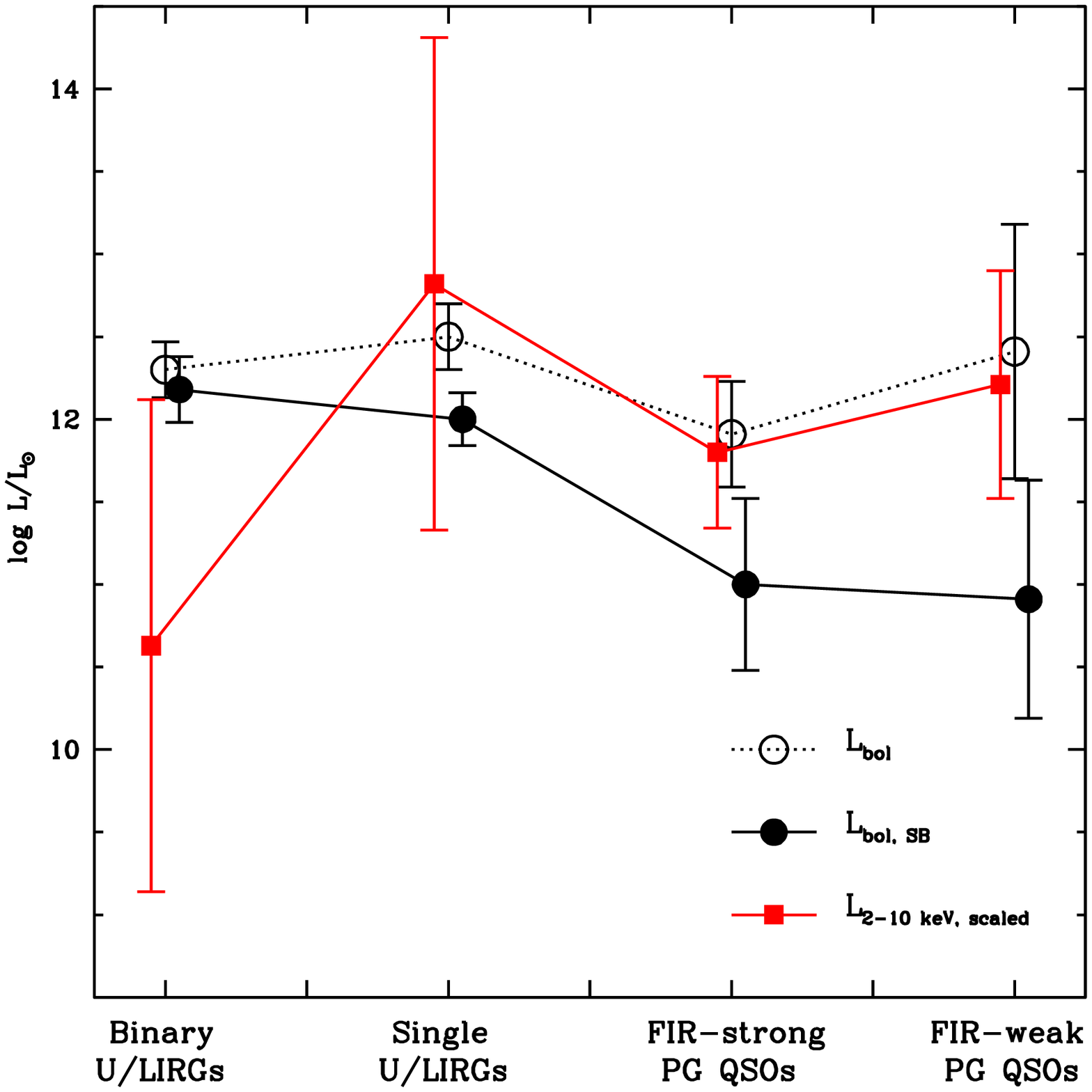}
\caption{The starburst bolometric luminosity is plotted as a function of
  merger stage.  On the horizontal axis, we have binary U/LIRGs
  \citep[interaction classes IIIab--IVa, phase D of][]{hopkins08},
  single-nucleus U/LIRGs (interaction classes IVb--V, phase D),
  FIR-strong PG~QSOs where star formation is still present but the
  black hole dominates the feedback process (the ``blowout'' stage,
  phase E), and the FIR-weak PG~QSOs where star formation has stopped
  (phase F).  The open circles trace the mean total bolometric
  luminosity of the objects from both the QUEST and RBGS samples, the
  filled circles show the mean {\em Spitzer}-derived starburst bolometric
  luminosity from V09a for the QUEST sample only, and the filled
  squares represent the mean absorption-corrected 2--10~keV luminosity
  for both the QUEST and RBGS samples multiplied by a factor of
  50 to be placed on the same scale as the other quantities.
  The error bars represent one standard deviation in each category of
  objects.  This figure suggests that: (1) the contribution of the
  starburst to the total bolometric luminosity decreases as the merger
  progresses, (2) there is essentially no difference in AGN power
  between the FIR-strong and FIR-weak PG~QSOs, (3) the growth of the
  AGN occurs most rapidly after coalescence, and (4) the 2--10~keV to
  bolometric luminosity correction for these AGNs is $\sim$50, the
  normalization factor between the infrared and X-ray values. The
  large error bars on the X-ray data may be attributed to poor
  absorption correction or a broad intrinsic distribution of the hard
  X-ray to bolometric ratios among pure AGNs. }
\label{fig:comp_model}
\end{figure}

\begin{figure}
\figurenum{14}
\epsscale{.8}
\centering
\plotone{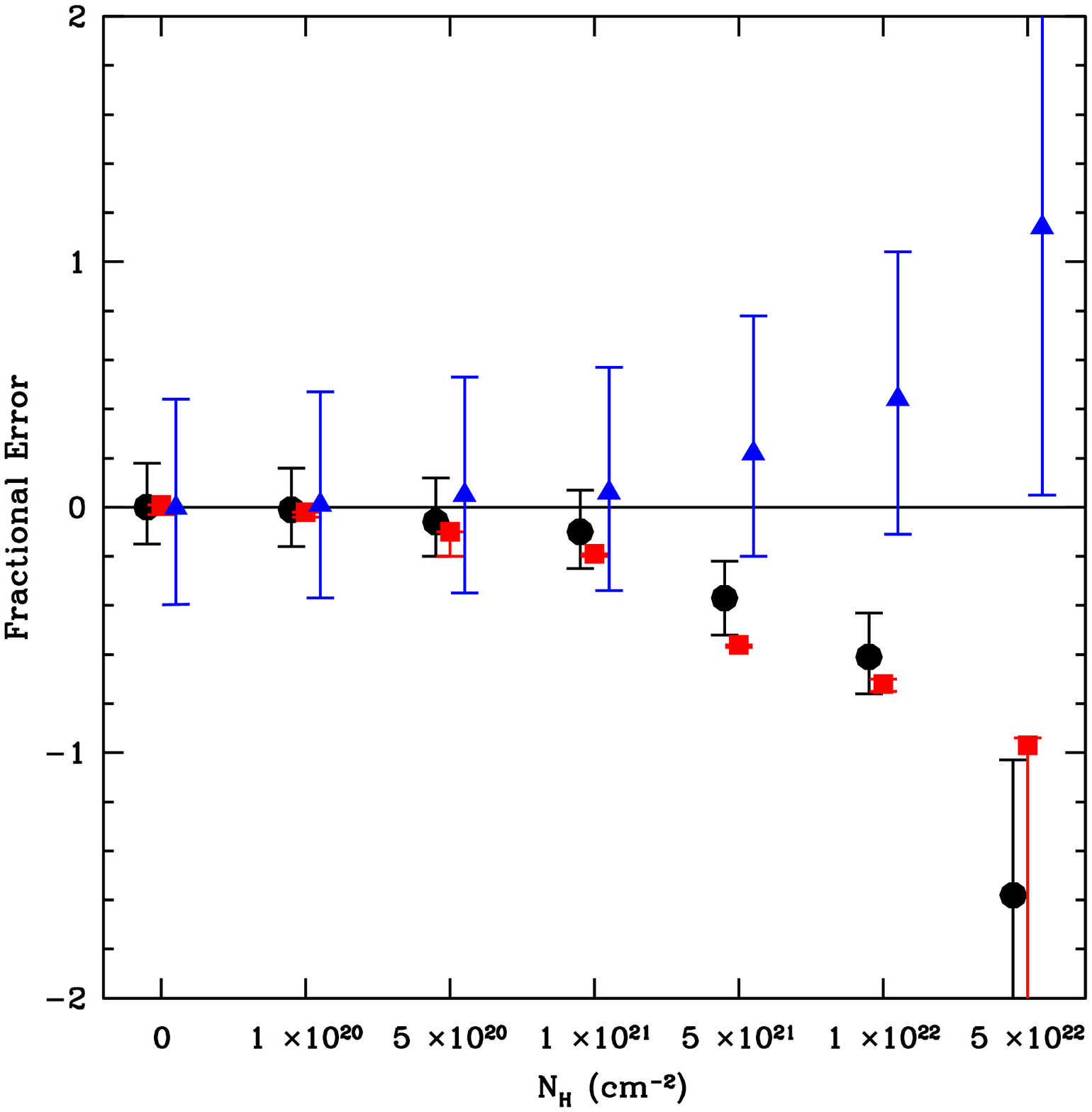}
\caption{The fractional errors of the HR method from the average of  1000 simulations.  The output photon index (black circles),  0.5--2~keV flux (red squares) and 2--10~keV flux (blue triangles),  both corrected for Galactic absorption from the simulated  \textit{Chandra} spectra, are plotted against the input intrinsic  column density.  It is clear from the plot that the photon index  derived from a single power law model deviates from the input value  of 1.8 when the source column density is $\gtrsim  10^{21}$~cm$^{-2}$, becoming flatter ($\Gamma_{\rm HR} \sim 0.71$ at  N$_H$ of $1 \times 10^{22}$~cm$^{-2}$).  The 0.5--2~keV flux, more  readily affected by absorption than the 2--10~keV flux, follows the  same trend as the photon index.  On the other hand, the hard-band  flux is stable up to N$_H \sim 5 \times 10^{21}$~cm$^{-2}$.  These  results of the simulations demonstrate that a flat spectrum is an  indication of obscuration in the source.}
\label{fig:hrsims}
\end{figure}

\begin{figure}
\figurenum{15}
\epsscale{.8}
\centering
\plotone{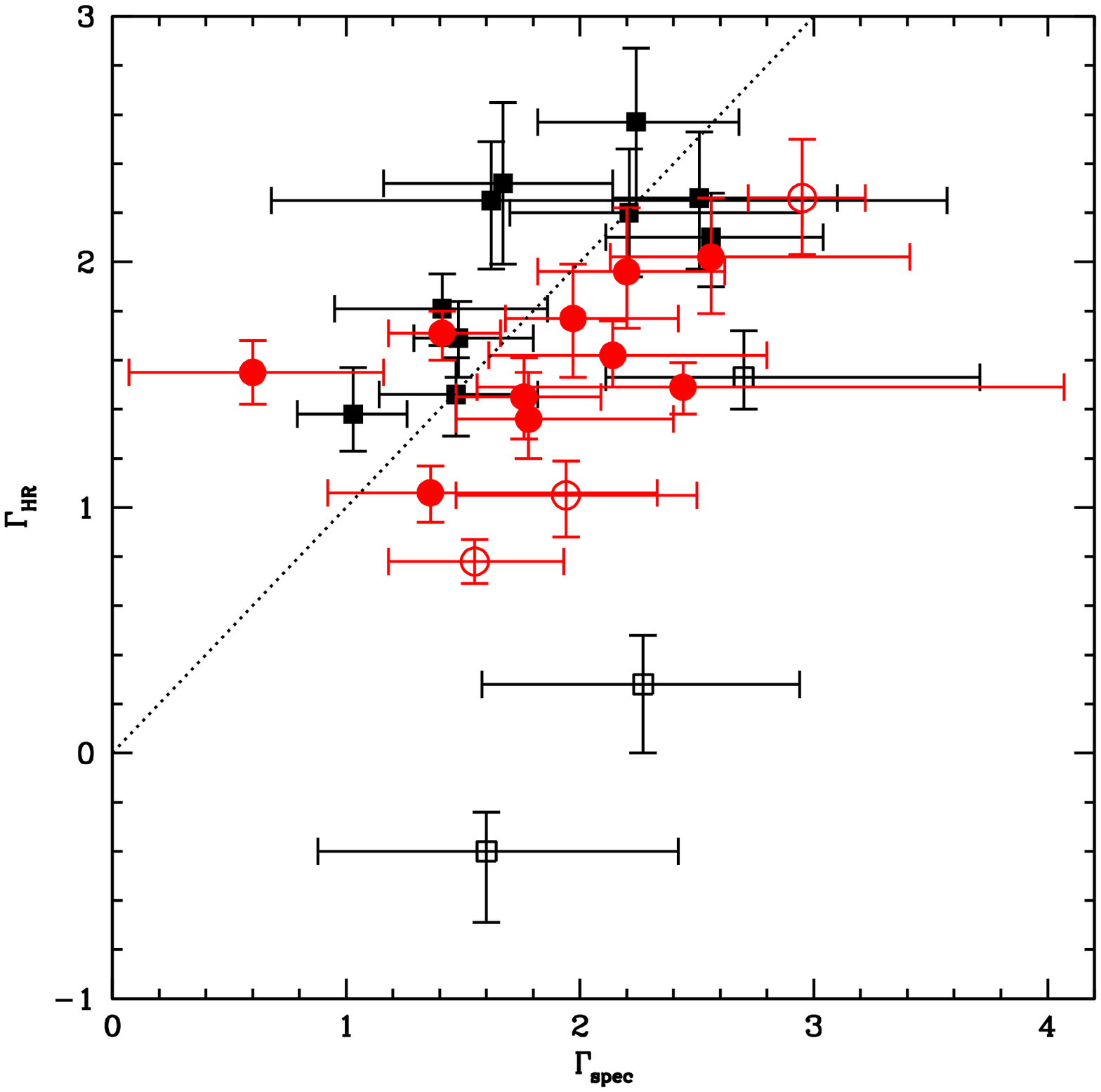}
\caption{A comparison of the values of $\Gamma$ derived from the
  traditional spectral fitting method (horizontal axis) and the HR
  method (vertical axis).  Included are the 13 objects from the QUEST
  sample (circles) that have enough counts for spectral fitting and
  also have been determined to have a power-law component in their
  spectra.  We add to these 13 more nuclei from the RBGS sample
  (squares).  The error bars are at 90\% confidence level.  The dotted
  line is the line of equality to help guide the eye.  For most of the
  objects (filled symbols), the values of $\Gamma$ derived from both
  methods are consistent with each other to within the errors.  For
  the six sources with open symbols, the hardness ratio method
  severely underestimates $\Gamma$.  All six objects have $N_H >
  10^{22}$~cm$^{-2}$.  This plot demonstrates that the hardness ratio
  method is a good estimator of the spectral properties of these faint
  sources as long as the column densities are $\lesssim
  10^{22}$~cm$^{-2}$, consistent with the results of our simulations
  (Table~\ref{tab:hrsims}).}
\label{fig:hrgamma}
\end{figure}

\begin{figure}
\figurenum{16}
\epsscale{1}
\centering
\plotone{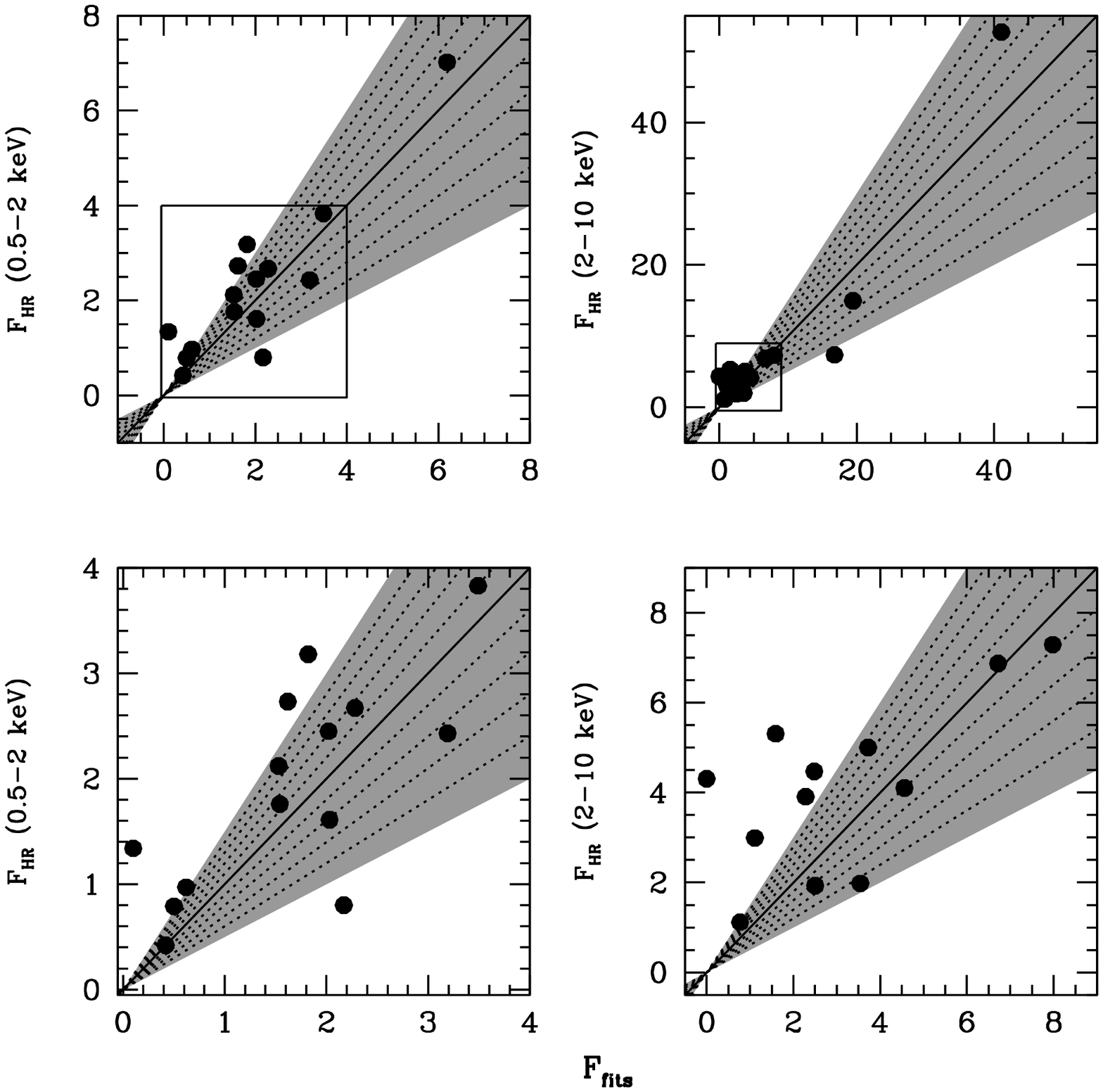}
\caption{A comparison of the 0.5--2~keV (left) and 2--10~keV (right) flux values of the U/LIRGs derived from spectral fitting (horizontal axis) and the HR method (vertical axis) for objects with enough counts for spectral fitting using c-stat.  The flux values are in units of $10^{-14}$~erg~s$^{-1}$~cm$^{-2}$.  The bottom figures are close-up views of the boxed regions in the top panels.  The solid line is a line of equality with each dotted line representing a 10\% deviation from the spectral fitting values.  Most of the HR values are within 50\% (the shaded regions) from the fitted values.  The HR method tends to overestimate the fluxes, especially at 2--10~keV, when obscuration is high.  The median values for $F_{HR}/F_{fits}$ are $\sim$1.2 and 1.3 for the soft and hard bands, respectively.   This is an indication that many of these objects are obscured (see Table~\ref{tab:hrsims}).}
\label{fig:hrcomp}
\end{figure}

\end{document}